\documentclass[preprint,12pt]{elsarticle}
\usepackage[utf8]{inputenc}

\usepackage{amssymb}
\usepackage{graphicx,indentfirst,amsmath,amsfonts,amssymb,amsthm,newlfont}
\usepackage{graphics}
\usepackage{subfigure}
\usepackage{graphicx}
\usepackage{epsfig}
\usepackage{xcolor}
\usepackage{float}

\begin{document}

\begin{frontmatter}

\title{Influence of nicotine and alcohol on sleep latency: the reward-attention circuit model}

\author[FIOCRUZ]{Karine Guimar{\~a}es}
\ead{karine.guimaraes@fiocruz.br}
\address[FIOCRUZ]{Oswaldo Cruz Foundation, Rio de Janeiro, Brazil}

\begin{abstract}
    {We previously developed a neurocomputational model called the reward-attention circuit, which has been used to investigate the separate effects of nicotine and alcohol on attention focus. The model is based on known circuits linking midbrain dopaminergic neurons to the thalamocortical loop and is sufficiently versatile to capture diverse phenomena. In this work, we used the model to study the influence of both nicotine and alcohol on sleep latency (time to fall asleep). In line with findings from other studies, our simulations suggest that nicotine can have a stimulating effect that increases sleep latency, promoting wakefulness, and, in contrast, alcohol can have a sedating effect that induces sleep. Together, our results suggest insights into alterations in mesothalamic dopamine activity and sleep. Also, they raise speculations about aspects of insomnia and low mesothalamic dopamine.
}
\end{abstract}

\end{frontmatter}

\section{Introduction}

Dopamine (DA) is a neurotransmitter known to regulate reward, learning, and movement~\cite{Klein2019}. However, its role as one of the neuromodulators responsible for wakefulness is emerging~\cite{Wisor2001, Qu2008, Dzirasa2006, Holst2014, Wisor2019}. Understanding the role of dopamine in the sleep-wake cycle is a question that naturally arises in this context. Substances that increase the level of dopamine in the brain, e.g. nicotine, increase the sensation of wakefulness~\cite{Boutrel2004}. DA also putatively participates in sleep regulation, as suggested by clinical evidence of sleep impairment in Parkinson's disease, a disease related to the degeneration of dopaminergic neurons~\cite{Dagher2009}. 

In addition to being the main integrative structure and relay of sensory and motor information to the cerebral cortex, the thalamus is also the first station at which incoming signals can be blocked by synaptic inhibition during sleep. This mechanism contributes to the transition from the waking to the sleeping state~\cite{Steriade1993, Sejnowski2000}.

Midbrain dopaminergic neurons play a key role in various brain functions such as voluntary movement and goal-directed behavior, as well as cognition, emotion, reward, motivation, working memory, associative learning, and decision-making~\cite{Grace2007, Redgrave2010, Schultz2010a, Gerfen2011, Chowdhury2013}. Consequently, changes in the dopaminergic midbrain system are related to
neurological and psychiatric diseases, such as Parkinson’s disease (PD), attention deficit hyperactivity disorder (ADHD), and addiction~\cite{Dagher2009, Gold2014}. Two areas in the midbrain are particularly associated with large populations of dopaminergic neurons: the ventral tegmental area (VTA) and the \emph {substantia nigra pars compacta} (SNpc)~\cite{Schultz2010b}.

Nicotine and alcohol increase activity in the dopaminergic system~\cite{Tiwari2020, DiChiara1997}. In particular, increased dopamine release in the \emph {nucleus accumbens} (NAcc) is associated with reward substance abuse. However, the molecular mechanism of addiction remains unclear as many brain systems are involved in it~\cite{Robison2011}.

Growing evidence points to a central role of DA in regulating sleep-wake states. Stimulants that enhance dopaminergic tone are among the most potent known wake-promoting substances~\cite{Boutrel2004} and their arousing effects are abolished in mice deficient in DA signaling~\cite{Wisor2001, Qu2008}. Partial DA depletion causes disturbances of REM sleep without affecting motor functions~\cite{Dzirasa2006}. Humans with reduced levels of DA reuptake transporter display altered slow-wave activity following sleep deprivation~\cite{Holst2014}.

Nicotine stimulates the release of sleep-regulating neurotransmitters, including dopa\-mine, leading to sleep disturbances such as latency in sleep initiation~\cite{Zhang2006}. Defined as the transition time from wakefulness to sleep, the sleep latency is greater among smokers due to the stimulating effects of nicotine which increases alertness~\cite{Silva2022} making it difficult to initiate sleep. The blood nicotine levels were shown to be similar after a single exposure to nicotine, in addition, second-hand smoke exposure is speculated to affect sleep through nicotine stimulating effect~\cite{Babanayagam2011}.

Blood alcohol levels may continue to rise for some time during sleep, depending on the timing of sleep onset concerning consumption~\cite{Rundell1972}. Alcohol initially acts as a sedative and reduced sleep latency is one of the commonly reported phenomena~\cite{Stone1980, Colrain2014, Angarita2016}.

Anatomical evidence accumulated over the past two decades has demonstrated that the thalamic reticular nucleus (TRN) has significant dopaminergic innervation originating from the SNpc~\cite{Freeman2001, Floran2004, Garcia2007, Monje2019}. The TRN plays a central role in the control of attention while ADHD is associated with genetic abnormalities of dopamine D4 receptors~\cite{LaHoste1996, Faraone2001}. There is suggestive evidence that abnormal dopaminergic transmission in TRN may generate at least some of the symptoms of ADHD~\cite{Floran2004, Wells2016}. 

The relatively new and unexplored thalamic dopaminergic pathway is the basis of the neurocomputational model called the reward-attention circuit that we have developed and used to investigate the separate effects of nicotine~\cite{Guimaraes2017} and alcohol~\cite{Guimaraes2018} on attention focus. Based on experimental evidence~\cite{Freeman2001, Floran2004, Garcia2007, Monje2019} and the efferent projections from NAcc to SNpc we model the interaction between the reward and thalamocortical circuits. The association between reward and thalamic circuits has already been addressed before through computational models~\cite{Humphries2006, Kumaravelu2016}. Our work, however, follows a particular direction since it highlights the importance of subcortical systems in attentional mechanisms and, in particular, the dopaminergic role in the thalamocortical circuit. Moreover, it assumes the coupling of the reward and thalamocortical circuits as a key element for understanding how nicotine and alcohol affect attentional focus. 

In this paper, we expand the reward-attention circuit model to address how nicotine and alcohol influence sleep. In its previous versions, the reward-attention model considered inattention symptoms in awake conditions~\cite{Guimaraes2017, Guimaraes2018}. Therefore, the previous studies assumed that the thalamus operated in its tonic state. However, the prevalent firing mode of thalamic neurons during sleep is the bursting mode~\cite{Steriade1993, Pace2002}, and the dynamics of ionic mechanisms under bursting are different from the dynamics under tonic mode.  

Here, we extend the previous studies and investigate the relationship between the nigral dopaminergic activity and the oscillatory state of neurons in the thalamic complex. Doing so, it becomes possible to widen the investigation to examine a possible mesothalamic DA activity contribution to sleep alterations in the presence of nicotine or alcohol. Despite experimental evidence, the neural mechanisms through which nicotine and alcohol act to modulate sleep are not yet fully elucidated.

Insomnia is defined as the symptom of difficulty in falling asleep, repeated awakenings with difficulty in returning to sleep, or sleep that is nonrestorative or poor in quality, often accompanied by the perception of short overall sleep duration. Moreover, insomnia is the most prevalent sleep disorder in people with Parkinson's disease~\cite{Loddo2017}, ADHD~\cite{Bijlenga2013}, and autism spectrum disorder (ASD)~\cite{Pekka2003}. 

In the previous versions of the reward-attention circuit model~\cite{Guimaraes2017, Guimaraes2018}, these disorders were treated as applications of the model to observe the focus of attention on individuals affected by them. In~\cite{Guimaraes2017} our results highlighted aspects of the lack of attention, particularly in ADHD. In \cite{Guimaraes2018} we simulated alcohol exposure in ASD disorder and the results suggested why people with ASD might relax and enhance attentional focus when exposed to alcohol. Since these findings are in line with clinical trials, this reinforces the validation of our model.

In the same way that we previously used attentional focus as a common symptom of PD, ADHD and ASD to evaluate potential applications of the model, here we will use insomnia as a common symptom of these disorders to explore additional applications of the model.

\section{The extended reward-attention model}

\begin{figure}
\begin{centering}
\resizebox{9cm}{!}{\includegraphics{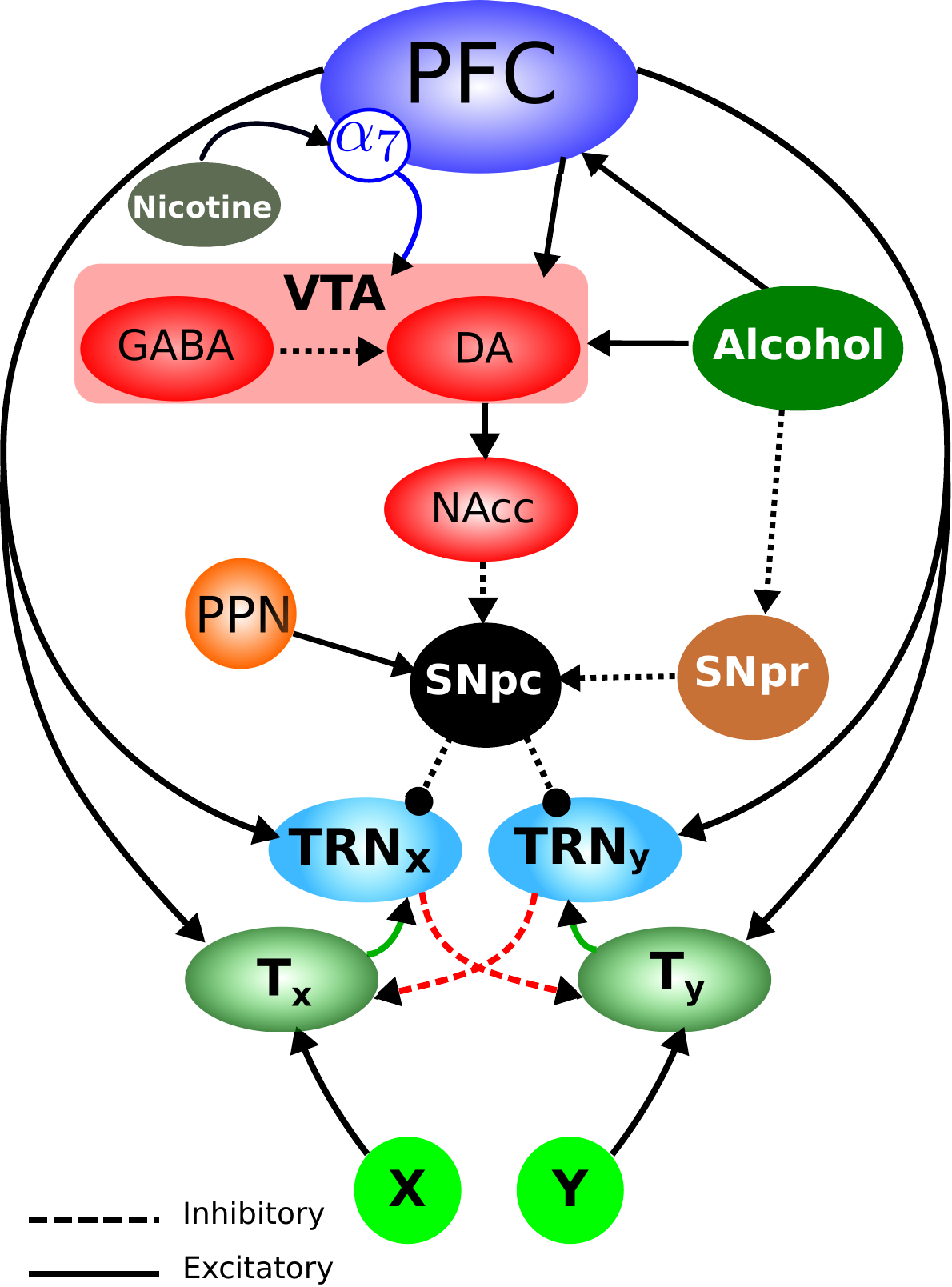}} 
\caption{\textbf{Network structure of the extended reward-attention model.} Brain regions are prefrontal cortex (PFC), ventral tegmental area (VTA), nucleus accumbens (NAcc), substantia nigra pars compacta (SNpc), substantia nigra pars reticulata (SNpr), thalamic reticular nucleus (TRN), and thalamic regions $T_x$ and $T_y$. Nicotine and alcohol act on the glutamatergic terminal in the prefrontal cortex (PFC). In VTA, dopaminergic neurons (DA) are inhibited by GABAergic interneurons. X and Y are sensory stimuli.} 
\label{fig:arquitetura_RAC}
\end{centering}
\end{figure}

The scheme of the extended reward-attention model is presented in Figure~\ref{fig:arquitetura_RAC}. The model consists of 10 nodes, representing brain areas, plus 4 nodes that represent stimuli $x$, $y$, nicotine, and alcohol. The nodes are interconnected by links that represent excitatory or inhibitory synapses. 

In our model, nicotine or alcohol is responsible for activating the reward circuit. As a consequence of the nicotinic stimulus, the pre-frontal cortex (PFC) excites VTA. In VTA, nicotine and the glutamate that is released by PFC stimulate GABAergic interneurons and DA neurons to release their respective neurotransmitters. Therefore, the GABAergic cells inhibit the DA cells. There are two types of nicotinic acetylcholine receptors (nAChR) that have been shown to mediate nicotine in the reward circuit, the $\alpha_7$ and $\alpha_4\beta_2$~\cite{Mansvelder2002, Srinivasan2011}. The $\alpha_4\beta_2$ receptors (not shown in Figure~\ref{fig:arquitetura_RAC}) desensitize faster than the $\alpha_7$ receptors and, as a consequence, the VTA GABAergic interneurons become inactive earlier while the VTA DA cells continue to be excited, both by nicotine and glutamate and by the lack of VTA GABAergic inhibition. The VTA DA neurons make excitatory synaptic connections with the NAcc GABAergic neurons that inhibit the DA neurons at SNpc. On the other hand, the stimuli $x$ and $y$ excite the neighboring thalamic areas $T_x$ and $T_y$, respectively. Excitatory projections from both $T_x$ and PFC stimulate $TRN_x$, which inhibits the thalamic area $T_y$. Similarly, $T_y$ and PFC excite $TRN_y$, which inhibits $T_x$. The DA neurons in SNpc modulate the activity of $TRN_x$ and $TRN_y$ by inhibiting this region.

On the other hand, the alcohol stimulus excites PFC and VTA and, as a consequence, PFC excites VTA. In VTA, alcohol and the glutamate that is released by PFC stimulate the GABAergic interneurons and the DA neurons to release their respective neurotransmitters. The rest of the circuit is the same as the one described above. These are the main aspects of the network. For more details on the original version of the model see~\cite{Guimaraes2017, Guimaraes2018}.

Before delving into the mathematical version of the model, it is worth paying attention to its mechanisms. Let us assume that sensory signals due to two external stimuli, $x$ and $y$, are conveyed through excitatory pathways to two neighboring thalamic regions $T_x$ and $T_y$, respectively. Once stimulated, $T_x$ and $T_y$ activate, respectively, $TRN_x$ and $TRN_y$ and send ascending glutamatergic projections to the cortex. Since we do not explicitly model the cortical area related to these sensory stimuli, only the excitatory projections from $T_x$ and $T_y$ to $TRN_x$ and $TRN_y$ are considered. However, we implicitly assume that PFC is eventually activated by sensory stimuli, and, explicitly, by nicotine or alcohol. Then, PFC sends descending excitatory glutamatergic projections to both $T_x$ and $T_y$ and also, via collateral axons, to $TRN_x$ and $TRN_y$.

Once activated, $TRN_x$, through its GABAergic projections, inhibits the neighboring thalamic region $T_y$. Reciprocally, $TRN_y$ inhibits $T_x$. In brief, the thalamocortical circuit activation by an external stimulus $x$ excites a central thalamic region $T_x$ and inhibits its neighborhood $T_y$, and vice-versa for an external stimulus $y$.

Additionally, $TRN_x$ and $TRN_y$ receive dopaminergic inhibitory projections from SNpc. A rise in the dopamine release by SNpc enhances the inhibition of $TRN_x$ and $TRN_y$ and leads to more active thalamic regions $T_x$ and $T_y$. Conversely, a reduction in the SNpc dopaminergic level strengthens the activity of $TRN_x$ and $TRN_y$ and diminishes the activity of $T_x$ and $T_y$.

DA neurons in VTA and SNpc are pacemakers, i.e., they are tonically active and display sustained firing in the absence of synaptic input~\cite{Guzman2009, Khaliq2010}. Besides, they can switch between tonic (pacemaker) and phasic (bursting) firing modes~\cite{Grace1991}. Electrophysiological recordings have demonstrated that these cells switch to firing in bursts of spikes due to salient stimuli and based on a prior history of reward experience (e.g. nicotine or alcohol)~\cite{Floresco2003}.

Bursts of spikes in the VTA DA neurons increase DA release in NAcc, which in turn promotes synaptic plasticity~\cite{Reynolds2001, Centonze2001} and drives reward-based reinforcement learning and goal-directed behavior~\cite{Schultz2013, Saunders2018}. Many studies have shown that burst firing activity is highly regulated by glutamatergic afferent inputs~\cite{Grace2007, Mansvelder2000, Bernier2011}.

It has been shown that activation of N-methyl-D-aspartate glutamate receptors (NMDARs) is necessary to enable the switch between tonic and phasic firing patterns~\cite{Sombers2009, Zweifel2009, Pignatelli2015}. The activation of NMDAR is voltage-dependent, triggering Ca$^{2+}$ influx which, in turn, increases the intensity and prolongs the duration of depolarization of the postsynaptic neuron. This phenomenon is called long-term potentiation (LTP) and is fundamental in complex neurobiological phenomena such as memory and learning~\cite{Cui2007, Foddai2003, Lisman1998}. Moreover, studies report that repeated \emph{in vivo} exposure to nicotine or alcohol causes enhancement of LTP of NMDAR-mediated transmission in VTA DA neurons~\cite{Mansvelder2000, Bernier2011}. Thus, both nicotine and alcohol can affect the pattern of activity in the reward circuit.

In our model, NAcc and SNpc provide a link between the reward and the thalamocortical circuits.  Besides the inhibitory GABAergic input from Nacc, SNpc also receives inhibitory projections from SNpr. There is experimental evidence that systemic administration of alcohol suppresses the firing of neurons in the SNpr~\cite{Mereu1985}.

Thalamic neurons are also able to fire in tonic and phasic modes~\cite{Steriade1993, Llinas2006}. When in a tonic state, these neurons respond linearly to input stimuli. In this way, they propagate information reliably from perceptual systems to the cerebral cortex, where more refined processing takes place. In our model, this mode of activity is crucial for the thalamocortical filtering of perceptual stimuli that allows attention focusing~\cite{Guimaraes2017, Guimaraes2018, Madureira2010}.

Conversely, when thalamic neurons are in phasic mode they are no longer reliable channels for sensory input to reach the cortex. The phasic mode underlies thalamic behavior during sleep~\cite{Steriade1993, Pace2002}. In this condition, environmental stimuli are not perceived consciously during wakefulness. The thalamic phasic mode also permeates epileptic episodes during which environmental information is not processed reliably~\cite{Jeanmonod1996}.

In summary, in our model the NAcc activity shapes the behavior of SNpc, thereby modulating the thalamocortical circuit. Our main hypothesis is that the influence of nicotine or alcohol is strong enough to bias the NAcc activation.

In previous works, we addressed inattention symptoms in wake subjects~\cite{Guimaraes2017, Guimaraes2018}. Because of this, we considered thalamic neurons firing in tonic mode. In this work, we study the relationship between the dopaminergic activity of SNpc and the bursting firing mode of thalamic neurons. This broadens the scope of applications of the model to explore the potential contributions of mesothalamic dopaminergic activity to sleep alterations when nicotine or alcohol is present.

\subsection*{Mathematical formulation}

We assume that each area comprises a population of neurons with homogeneous behavior, so we model each area by a single neuron. The currents that operate in every single neuron are based on neurophysiological and neurochemical aspects of the corresponding neuronal population.
This approach has been previously used elsewhere~\cite{Guimaraes2017, Guimaraes2018, Carvalho1994, Madureira2010, Carvalho1995}. For a discussion of homogenization in neuroscience, see, e.g.,~\cite {Cioranescu2000, Bressloff2014}.

To facilitate the computational implementation, the spiking of neurons is described by the leaky integrate-and-fire (LIF) model~\cite{Burkitt2006}. Thus, to each $n = 1,\dots, \mathcal{N}$ of the $\mathcal{N}$ neurons in the network, the equation that describes the evolution of the membrane potential is
\begin{equation}\label{eq:voltage}
C_n \dfrac{\mathrm{d}V_n}{\mathrm{d}t} =  \sum \limits_{j = 1}^{J_n} I^j_n + I_\mathrm{ext} \qquad \text{for }t \in (0, T] \; \text{and} \; V_n < \theta_{\text{Na}},
\end{equation}
where $C_n$ is the membrane capacitance, $J_n$ is the number of ionic currents in the membrane of the $n$th neuron, $I_\mathrm{ext}$ is the external current, and $\theta_{\text{Na}}$ is a fixed constant (the threshold). The initial condition is $V_n(0) =  V_n^0$ and $T$ is the total simulation time. If $V_n \geqslant \theta_{\text{Na}}$ the membrane potential of the neuron is reset instantaneously to the resting potential and a spike occurs. 

We assume that all neurons have three ionic currents, associated with potassium $(I^K_n)$, sodium $(I^{\text{Na}}_n)$, and leakage $(I^L_n)$. The potassium and leakage currents are modeled by 
\begin{equation}\label{eq:ionic_current}
    I^j_n = g_j(V_n) (E_j-V_n), 
\end{equation} 
where $g_j$ is the conductance associated to the $j$th current and $E_j$ is the corresponding Nernst potential. The leakage conductance is constant but the potassium conduction is not (see below). The sodium current is responsible for the spike and is given by the Heaviside function $\Theta:\mathbb{R} \rightarrow \{0,1\}$, defined by

\begin{equation}\label{eq:FuncaoHeaviside}
\Theta(V_n - \theta_{\text{Na}})=
\begin{cases}
1& \text{if } (V_n - \theta_{\text{Na}}) \geqslant 0 \\
0& \text{if } (V_n - \theta_{\text{Na}}) < 0,
\end{cases}
\end{equation}
where $\theta_{\text{Na}}$ (the threshold) represents the depolarizing effect of the sodium current.

After a spike, during the repolarization phase, the potassium conductance increases rapidly, driving the membrane potential towards the Nernst potential of potassium $E_\text{K}$,
\begin{equation}
\dfrac{\mathrm{d}g_{\text{K}}}{\mathrm{d}t} = \dfrac{\beta_{\text{K}}\Theta(V_n - \theta_{\text{Na}}) - g_{\text{K}}}{\tau_{\text{K}}} \qquad \text{for }t \in (0, T],
\label{eq:potassium_conductance}
\end{equation}
where $g_{\text{K}}(0) = g_{\text{K}}^0$ is the initial condition,  $\beta_{\text{K}}$ is the variation rate of $g_{\text{K}}$, and $\tau_{\text{K}}$ is the time constant of the potassium conductance. 

The current $I_\mathrm{ext}$ represents the sum of the total synaptic currents and external currents that are applied due to the eventual presence of nicotine or alcohol. The sum of excitatory and inhibitory synapses acting on each neuron in the network is given by

\begin{equation} \label{eq:synaptic_current}
I_{\text{syn}}(t) = g_{{\text{E}}}(t) (E_{\text{E}} - V_n) + g_{\text{I}}(t) (E_{\text{I}} - V_n) ,
\end{equation}
where $g_{{\text{E}}}$ and $g_{\text{I}}$ are, respectively, the total conductances of all excitatory and inhibitory synapses on the neuron, and $E_{\text{E}}$, $E_{\text{I}}$ are their respective reversal potentials.

The time variation of the synaptic conductances due to a sequence of synaptic arrivals is described by a sum of alpha functions~\cite{Ermentrout2010}
\begin{equation}\label{eq:synaptic_conductance} 
g_{\text{syn}}(t) = \hat{g}_{\text{syn}}\sum_{j}(t - t_j)\exp \biggl(-\frac {t - t_j}{t_p}\biggr) \Theta(t-t_j),
\end{equation}
where the subscript syn $\in$ {E, I}, the times t$_j$ denote the firing times of presynaptic neuron $j$, and the constant $\hat{g}_{\text{syn}}$ is the maximal synaptic conductance. The parameter $t_p$ is the peak time of the alpha function, and it assumes values $t_{pe}$ and $t_{pi}$ for excitatory and inhibitory synapses, respectively.  

In the model, we assume that nicotine or alcohol (at different times) are responsible for activating the reward circuit. Both stimulate the PFC neuron as external currents. 

The rate of change of the number of active nicotinic receptors is given by
\begin{equation}
\dfrac{\mathrm{d}\alpha_7}{\mathrm{d}t} = k_1 \alpha_4\beta_2 [C_{10}H_{14}N_2] - k_2\alpha_7 \quad\text{for }t \in (0, T], 
\label{eq:alpha7}
\end{equation}
where $\alpha_7(0) = \alpha_7^{0}$ and $\alpha_7^{0}$, $\alpha_4\beta_2$, $k_1$ and $k_2$ are constants and $[C_{10}H_{14}N_2]:(0,T] \rightarrow \mathbb{R}$ is the nicotine and is given by the solution of the following differential equation 
\begin{equation}
\dfrac{\mathrm{d}[C_{10}H_{14}N_2]}{dt} = - \kappa [C_{10}H_{14}N_2] \quad\text{for }t \in (0, T], 
\label{eq:nicotine}
\end{equation}
where $[C_{10}H_{14}N_2](0) = [C_{10}H_{14}N_2]^0$ and $\kappa$ and $[C_{10}H_{14}N_2]^0$ are constants.

Alcohol, in turn, activates the PFC and VTA GABAergic interneurons, which are afferent to VTA dopaminergic neurons. The effect of alcohol is represented by $C_2H_6O:(0,T] \rightarrow \mathbb{R}$, which is solution of the differential equation 
\begin{equation}
\dfrac{\mathrm{d}[C_2H_6O]}{dt} = - \lambda [C_2H_6O] \quad\text{for }t \in (0, T],
\label{eq:alc}
\end{equation}
where $[C_2H_6O](0) = [C_2H_6O]^0$ and $\lambda$ and $[C_2H_6O]^0$ are constants.


In resting conditions, VTA DA neurons exhibit a tonic firing pattern~\cite{Koyama2005,Margolis2006}. In the model, the pacemaker current in the VTA DA neuron is given by $I_{\text{pm}} = g_{\text{pm}} (V_n - E_{\text{pm}})$, with $g_{\text{pm}}$ and $E_{\text{pm}}$ constants.

Alcohol increases the spontaneous firing frequency of VTA DA neurons in a concentra\-tion-dependent manner~\cite{Brodie1990, Okamoto2006}. To model this effect, when $[C_2H_6O]^0\ne0$ we add $[C_2H_6O]$ given by~\eqref{eq:alc} to $g_{\text{pm}}$ in the VTA DA neuron. 

To allow the VTA DA neuron to fire in burst mode, we added to this neuron an NMDAR conductance and a calcium-activated potassium current $(I_{\text{AHP}})$. The conductance of the NMDAR-mediated current is given by $g_{\text{\tiny{NMDA}}} = \bar{g}_{\text{\tiny{NMDA}}} h(t)B(V_{\text{VTA}})$, 
where $V_{\text{VTA}}$ is the voltage of the VTA DA neuron, and $h(t)$ is the fraction of open channels and satisfies
\begin{equation}
\dfrac{\mathrm{d}h}{\mathrm{d}t}=a_{\text{r}}(1-h)\,\dfrac{\mathcal{T}_{\text{max}}}{1 + e^{-(V_{\text{PFC}} - V_{\mathcal{T}})/k_\text{p}}}-a_{\text{d}}h\quad\text{for }t\in(0,T],
\label{eq:NMDA_concentration}
\end{equation}
where $h(0)=h^0$, and $a_{\text{r}} = 0.072$ mM$^{-1}$ms$^{-1}$ and $a_{\text{d}} = 0.0066$ ms$^{-1}$ are the rates of increase and decay of $h$, respectively.  The parameter $\mathcal{T}_{\text{max}} \in \mathbb{R}$ is the maximum concentration of neurotransmitters in the synaptic cleft, $V_{\text{PFC}}$ is the voltage of the PFC neuron, $k_\text{p} =-5$mV is the decay parameter for neurotransmitters, and $V_{\mathcal{T}} = - 10$ mV is the value at which the function is activated.

The NMDAR is subject to voltage-dependent blocking caused by magnesium ions (Mg$^{2+}$). When the postsynaptic neuron is depolarized, the magnesium ions are removed and current can flow through the receptor. This mechanism was described in the model by the term $B(V_{\text{VTA}})$, which gives the probability of unblocking as a function of $V_{\text{VTA}}$~\cite{Jahr1990},
\begin{equation}
B(V_{\text{VTA}}) = \frac{1}{1 + e^{-(V_{\text{VTA}} - V_{\text{T}})/16.13}}, 
\label{eq:MagnesiumBlock}
\end{equation}
where $V_{\text{T}} = 16.13\ln\left(\tfrac{[Mg^{++}]}{3.57}\right)$ is the half activation.

As a consequence of NMDAR activation, Ca$^{2+}$ channels open. The calcium conductance $g_{\text{c}}$ is proportional to the intracellular calcium concentration $[\text{Ca}]_i$, $g_{\text{c}} = \bar{g}_{\text{c}}[\text{Ca}]_i$, where $\bar{g}_{\text{c}}$ is a constant rate. The calcium concentration in the cell is described by
\begin{equation}
\dfrac{\mathrm{d}[\text{Ca}]_i}{\mathrm{d}t} = \dfrac{\beta_{[\text{Ca}]_i}\,\Theta(V_{\text{VTA}}-\theta_{\text{Na}}) - [\text{Ca}]_i}{\tau_{[\text{Ca}]}} \quad\text{for }  t \in (0, T], 
\label{eq:Ca_concentration}
\end{equation}
where ${[\text{Ca}]}(0) = [\text{Ca}]^0$, $\beta_{[\text{Ca}]}$ and $\tau_{[\text{Ca}]}$ are constants and represent the initial condition, the rate of intracellular calcium concentration variation and a time constant, respectively. The Heaviside function $\Theta(V_{\text{VTA}}-\theta_{\text{Na}})$ raises the calcium concentration whenever there is a neuronal spike. When the intracellular calcium concentration reaches a threshold value $\theta_{[\text{Ca}]}$, the $K^{+}$ ion channels of the hyperpolarizing current $I_{\text{AHP}}$ open and the conductance $g_{\text{AHP}}$ increases at rate $\beta_{\text{AHP}}$. This is represented by the equation,
\begin{equation*}
\dfrac{\mathrm{d}g_{\text{AHP}}}{\mathrm{d}t} = \dfrac{\beta_{\text{AHP}}\,\Theta([\text{Ca}]_i-\theta_{[\text{Ca}]})- g_{\text{AHP}}}{\tau_{\text{AHP}}} \quad\text{for }t \in (0, T],
\label{eq:gahp_condutance}
\end{equation*}
where ${g_{\text{AHP}}}(0) = g_{\text{AHP}}^0$ is the initial condition and $\tau_{\text{AHP}}$ is a time constant.
 
VTA DA neurons project to NAcc, the final site currently associated with the pleasure sensation triggered by nicotine or alcohol. From there, NAcc sends GABAergic input to SNpc~\cite{Wise2002}. In addition, within the \emph{substantia nigra} the SNpr provides GABAergic afferents to SNpc DA neurons.

In the part of the model that describes the thalamocortical circuit, based on experimental evidence~\cite{Freeman2001, Floran2004}, we assume that the SNPc neuron makes inhibitory dopaminergic synapses on the TRN neurons. This dopaminergic action is mediated by calcium-dependent potassium channels~\cite{Floran2004}. The activation of these channels causes potassium to leave the cell and decreases its firing rate. The consequence of this is a reduction in the inhibition of the thalamic neurons $T_x$ and $T_y$ by TRN cells~\cite{Madureira2010}.

The conductance of the calcium-dependent potassium channel in the TRN neurons is described by $g_{\text{kc}} = \hat{g}_{\text{c}} \,D_4^* \,S([\text{Ca}_i])$, where $\hat{g}_\text{c}$ is a proportionality constant, $D_4^*$ is the time-varying activation of dopamine $D_4$ receptors and $S([\text{Ca}_i])$ is a sigmoid function describing the increase in the intracellular calcium concentration after a presynaptic spike. The $D_4$ activation is modeled by a sum of alpha functions,
\begin{equation} 
D_4^*(t)=\hat{g}_{\text{d4}} \sum_{j}  (t - t_j)\exp \biggl(-\dfrac {t - t_j}{t_{\text{pd}}}\biggr) \Theta(t-t_j),
\label{eq:DA_receptor}
\end{equation}
where $\hat{g}_{\text{d4}}$ is a proportionality constant, the times $t_j$ are the spike times of the presynaptic neuron, and $t_{\text{pd}}$ stands for the peak time of the alpha function.

The sigmoid function $S([Ca]_i)$ is given by
\begin{equation}
S([\text{Ca}_i])= \dfrac{1}{1+\exp(-\alpha [\text{Ca}_i])},
\label{eq:Sigmoide}
\end{equation}
where the constant $\alpha$ controls the slope of $S$. 

The intracellular calcium concentration in the TRN neurons is described as in equation~\eqref{eq:Ca_concentration}, with $V_{\text{VTA}}$ replaced by $V_{\text{TRN}_x}$ or $V_{\text{TRN}_y}$. Thus, as the cell voltage grows past a threshold the conductance $g_{\text{kc}}$ increases, and the cell is inhibited by the dopaminergic input. 

The bursting activity of the thalamic neurons is describhed by the same NMDAR- and calcium-mediated mechanism described by equations~(\ref{eq:NMDA_concentration}), (\ref{eq:MagnesiumBlock}) and~(\ref{eq:Ca_concentration}).

The differential equations were numerically solved using Euler's method. All the relevant model parameters with their respective values are given in Table~\ref{tab:parameters} in Appendix A. These parameters were fitted to each neuron separately by comparison to the electrophysiological behavior of typical neurons from the brain regions considered in the model~\cite{Guimaraes2017, Guimaraes2018}. 

\section{Results}



During sleep, there is a cyclic occurrence of rapid eye movement (REM) and non-REM (NREM) phases~\cite{Brown2012, Luppi2019}. NREM sleep is divided into three stages (N1, N2, and N3), and the higher the stage, the deeper the sleep and the lower the frequency band of the oscillations. Stages N2 and N3 correspond to the so-called slow wave sleep (SWS) stage~\cite{Feld2015}. 

In this work, we are interested in the beginning of sleep, i.e. stage N1 or NREM1, when the transition from wake to deep sleep is effective. We will consider that before this stage the thalamic neurons fire tonically during wake, and at its onset, they change to phasic or bursting firing mode.

\subsection{Reference network}
In this section, we present simulation results relative to what we call a `reference network'. This corresponds to a `healthy' individual who has not been exposed to nicotine ($[C_{10}H_{14}N_2]^0=0$ in equation~\eqref{eq:nicotine}) or alcohol ($C_2H_6O^0=0$ in equation~\eqref{eq:alc}), and has no associated sleep disorders. The importance of determining the behavior of the reference network is that it sets a benchmark against which other types of behavior can be compared~\cite{Carvalho1994, Madureira2010, Guimaraes2017, Guimaraes2018}.

\begin{figure}[H]
\centering
{\includegraphics[width=0.495\textwidth]{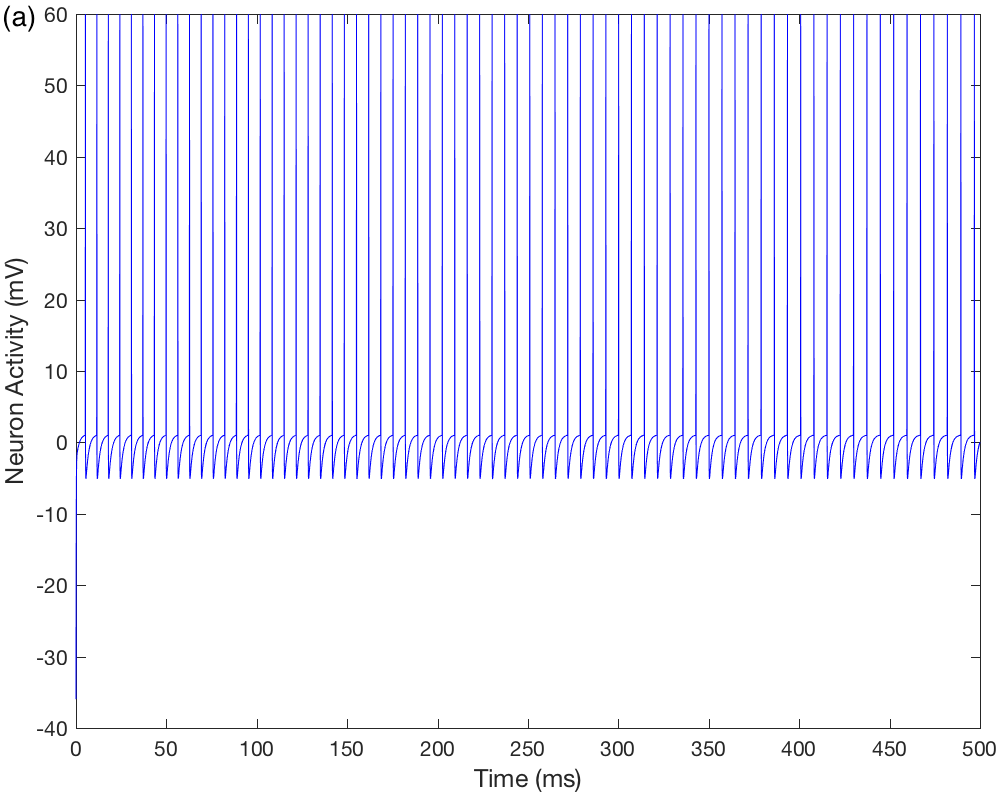}\label{fig:DA_basal}}
{\includegraphics[width=0.495\textwidth]{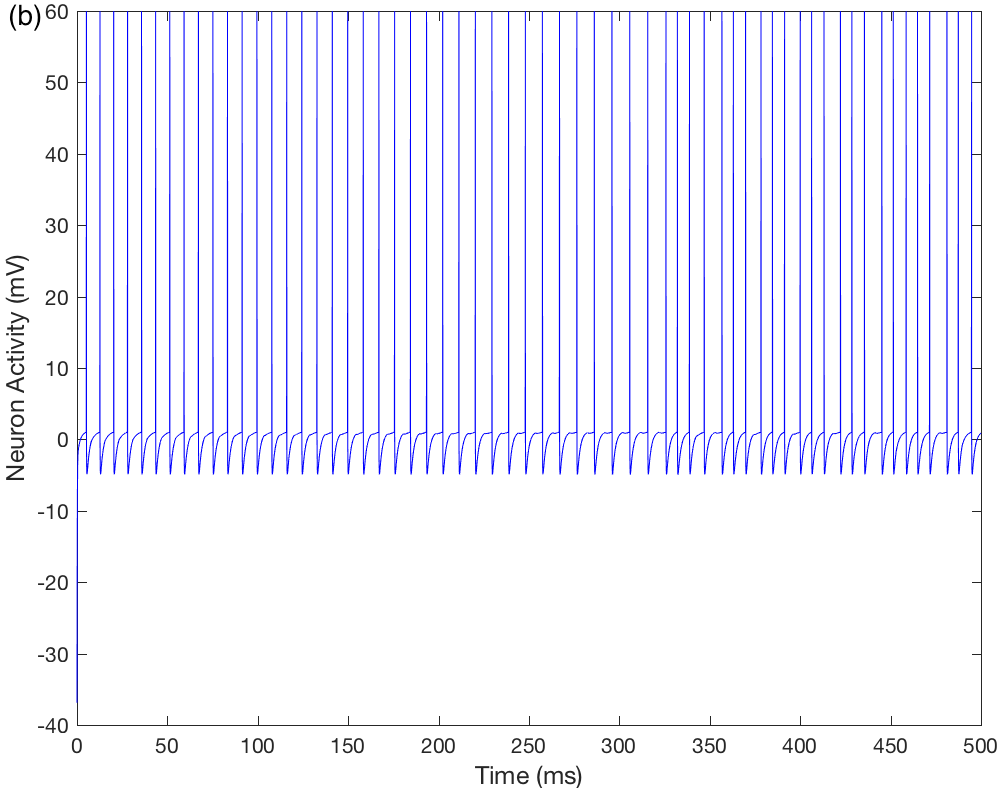}\label{fig:SN_basal}}
{\includegraphics[width=0.495\textwidth]{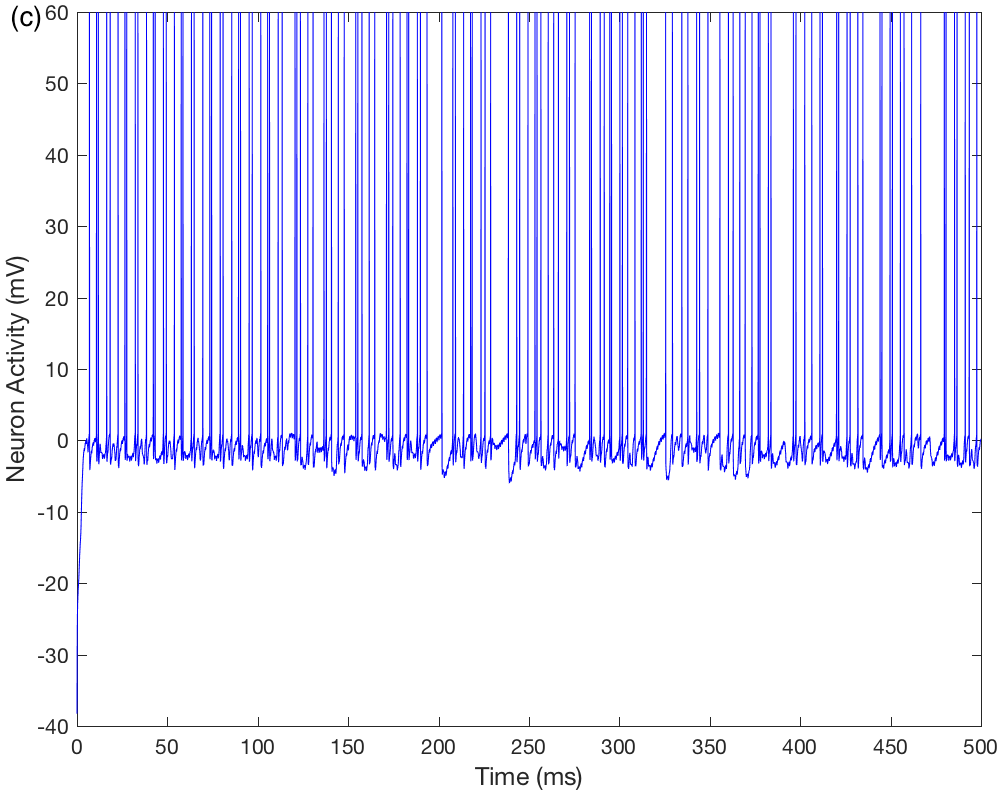}\label{fig:NRTx_basal}}
{\includegraphics[width=0.495\textwidth]{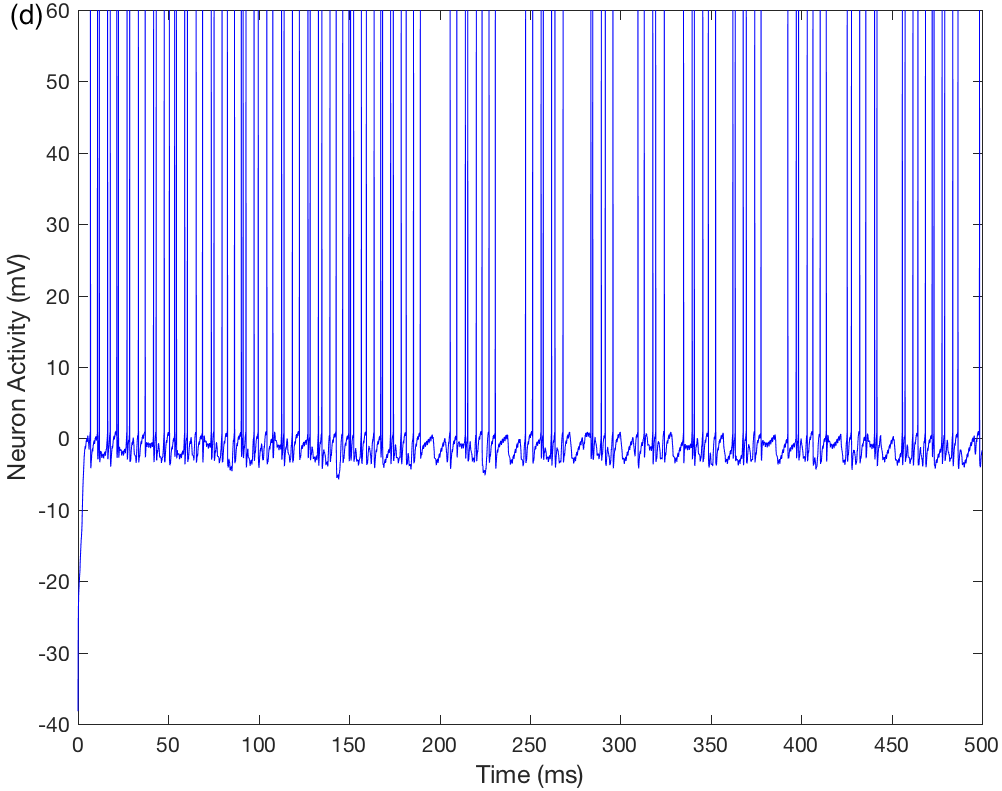}\label{fig:NRTy_basal}}
{\includegraphics[width=0.495\textwidth]{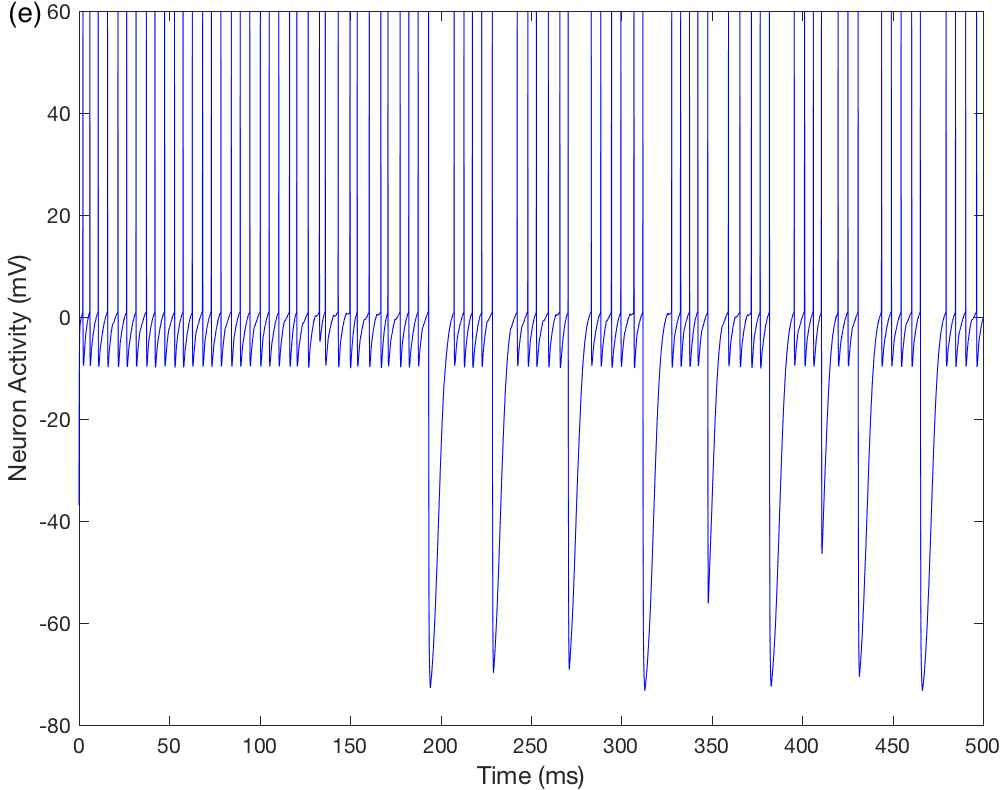}\label{fig:Tx_basal}}
{\includegraphics[width=0.495\textwidth]{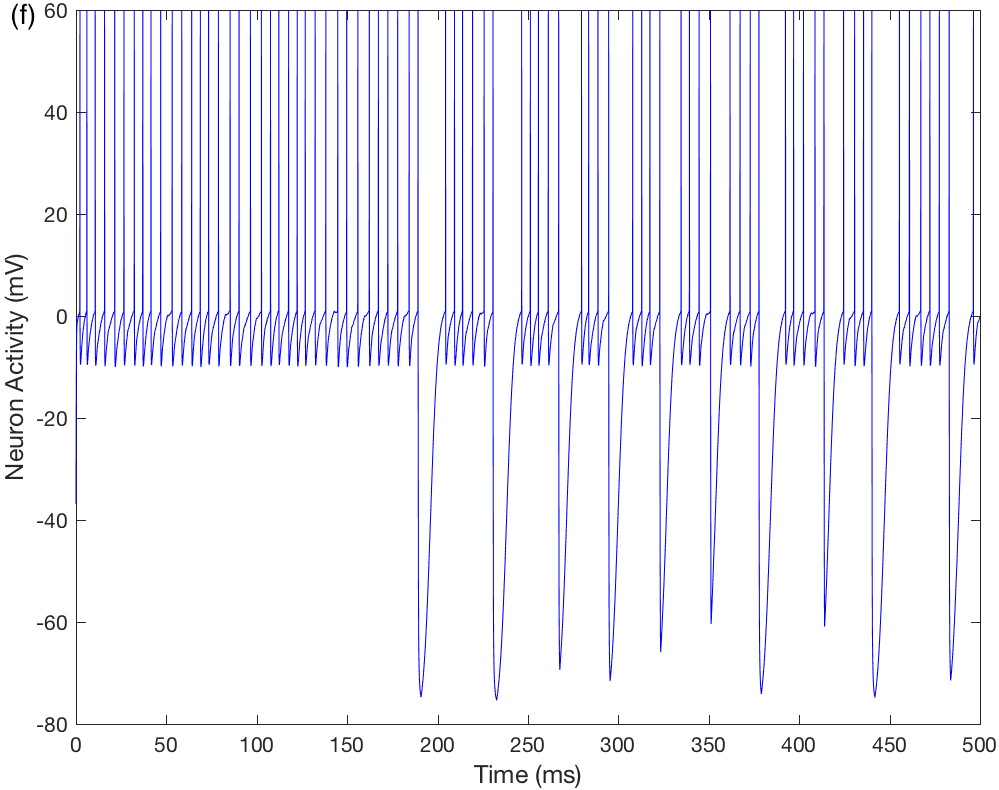}\label{fig:Ty_basal}}
\caption{\textbf{Behavior of network neurons for the reference network.} (a) VTA DA; (b) SNpc; (c) $TRN_x$;  (d) $TRN_y$; (e) $T_x$; (f) $T_y$.}\label{fig:SleepNormal}
\end{figure}

The results of the \emph{in silico} experiments are plotted here in graphs that show the neuronal voltages (or activities) over time, allowing visual observation of the spikes. In Figures \ref{fig:SleepNormal}a--f, we show the voltage behavior of the VTA DA, SNpc, $TRN_{x}$, $TRN_y$ $T_x $ and $T_y$ neurons.

We consider that the reward system remains almost inactive in the absence of nicotine or alcohol. Then, in the reward circuit, the VTA DA neuron displays a pacemaker (tonic) activity (Figure~\ref{fig:SleepNormal}a) providing the NAcc neuron with a baseline level of dopamine. The PFC and the VTA GABAergic neurons do not spike in this case.

The SNpc activity is modulated by the excitatory PPN and inhibitory SNpr synapses. Via its dopaminergic inhibitory synapses, the SNpc neuron keeps $TRN_{x}$ and $TRN_y$ under inhibitory control, modulating the degree to which $TRN_{x}$ and $TRN_y$ inhibit $T_y$ and $T_x$, respectively.

During a 500 milliseconds simulation, the PPN and SNpr spikes, the excitatory inputs $x$, $y$, and the PFC projection on the thalamocortical circuit are represented by time-periodic sequences of 1 spike per millisecond. These results will be used as a reference for simulations with presentations of nicotine and alcohol, and and with application insomnia. 

\subsection{Action of nicotine}

Our next simulation adds nicotine to the system by imposing $[C_{10}H_{14}N_2]^0 = 300$ nM in equation~\eqref{eq:nicotine}, which corresponds to the nicotine dose in a cigarette. All other parameters are the same as in the reference network. 

Due to the presence of nicotine, both the PFC and the VTA GABAergic neurons become active and the VTA DA neuron starts to receive glutamatergic and GABAergic stimuli. The VTA DA neuron is equipped with $\alpha_7$ nicotinic receptors, but the VTA GABA neuron is not. This models the experimental evidence that nicotinic receptors in VTA GABAergic neurons are of the $\alpha_4\beta_2$ type, which desensitizes faster, while the nicotinic receptors in VTA DA neurons are of the $\alpha_7$ type. As a consequence, the activity of the VTA GABAergic neuron decays while the VTA DA neuron continues to receive excitatory inputs and LTP plausibly occurs.

\begin{figure}[H]
\centering
{\includegraphics[width=0.495\textwidth]{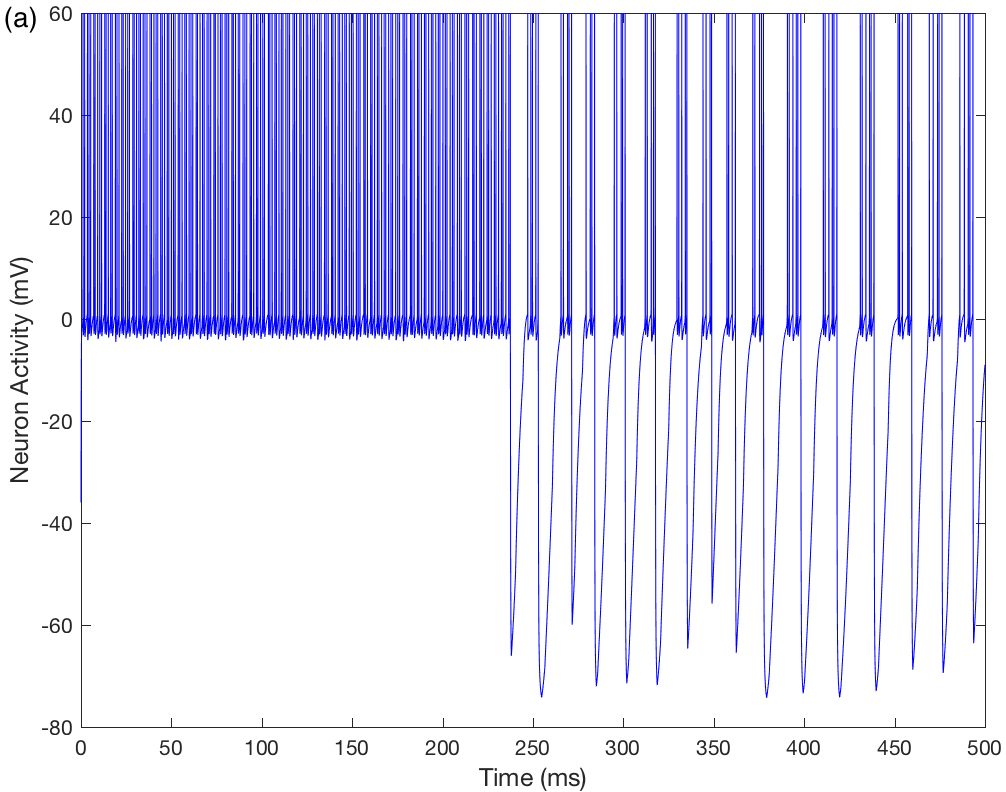}\label{fig:VTA_nic}}
{\includegraphics[width=0.495\textwidth]{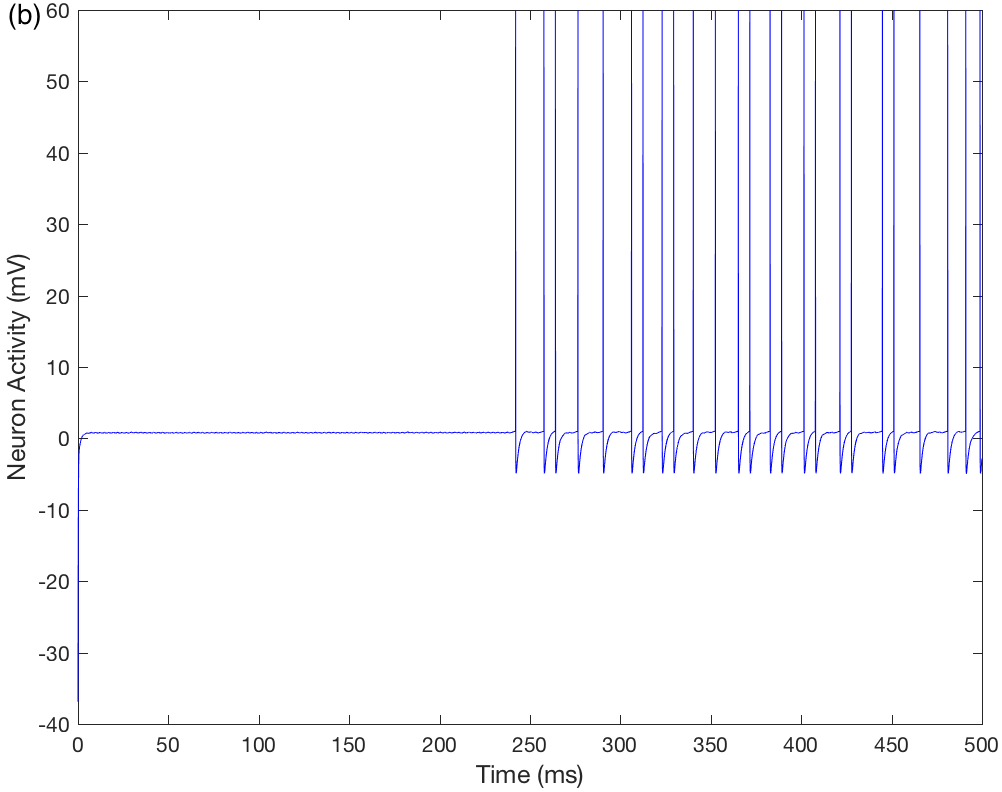}\label{fig:SN_nic}}
{\includegraphics[width=0.495\textwidth]{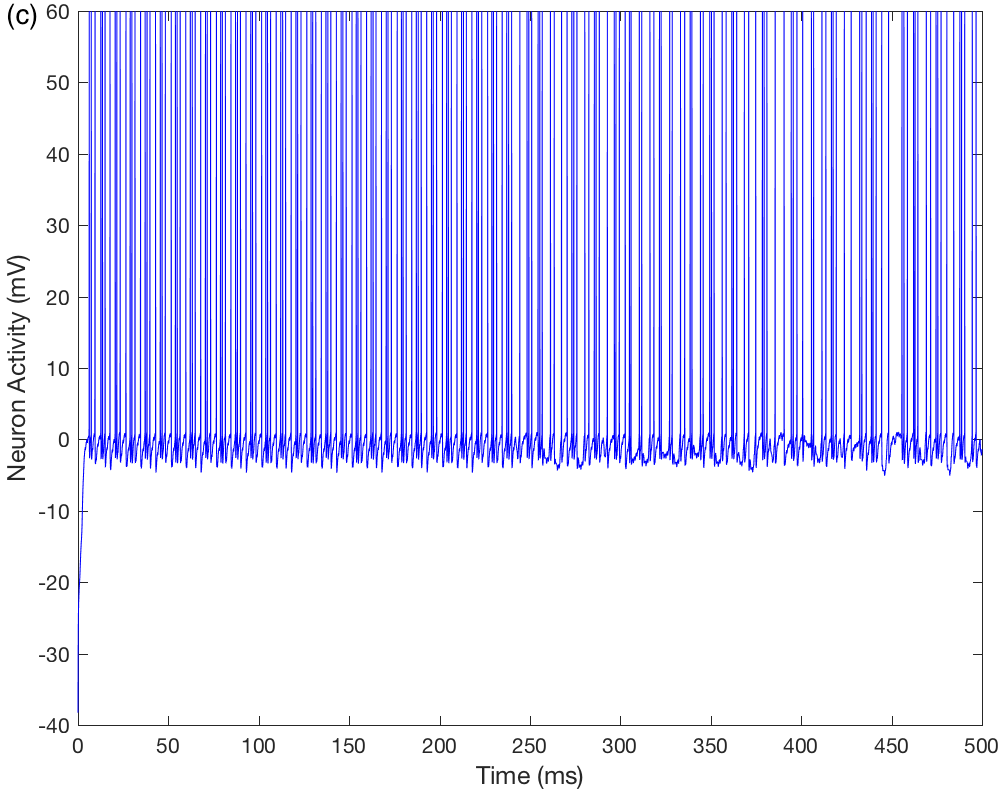}\label{fig:NRTx_nic}}
{\includegraphics[width=0.495\textwidth]{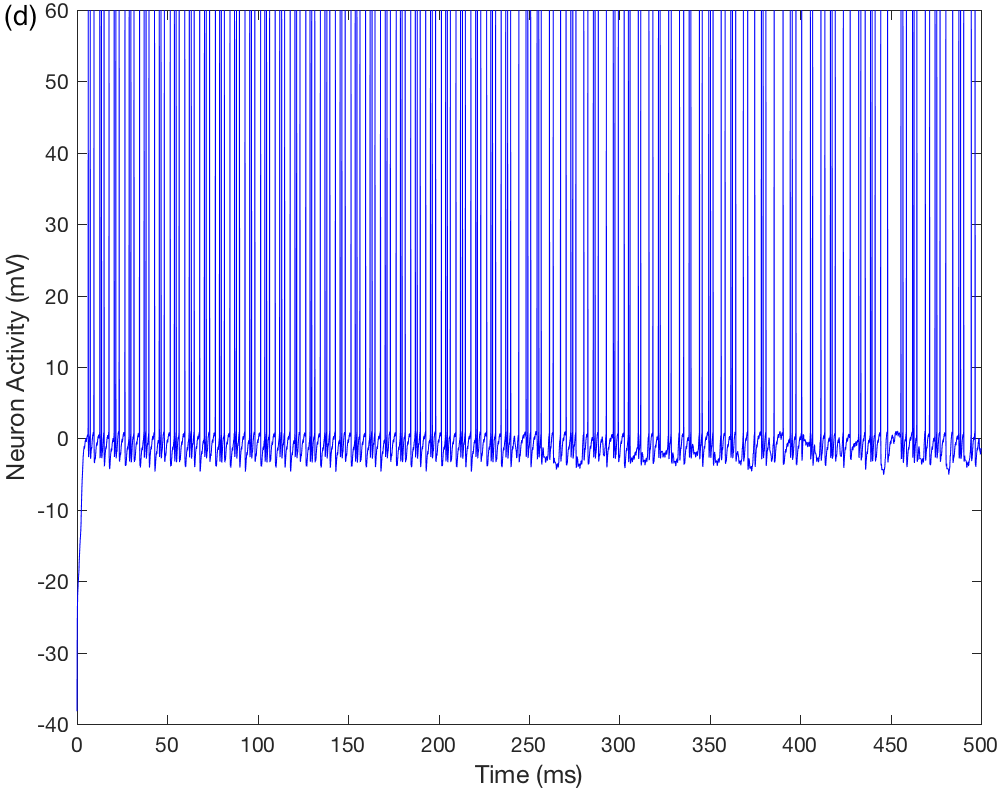}\label{fig:NRTy_nic}}
{\includegraphics[width=0.495\textwidth]{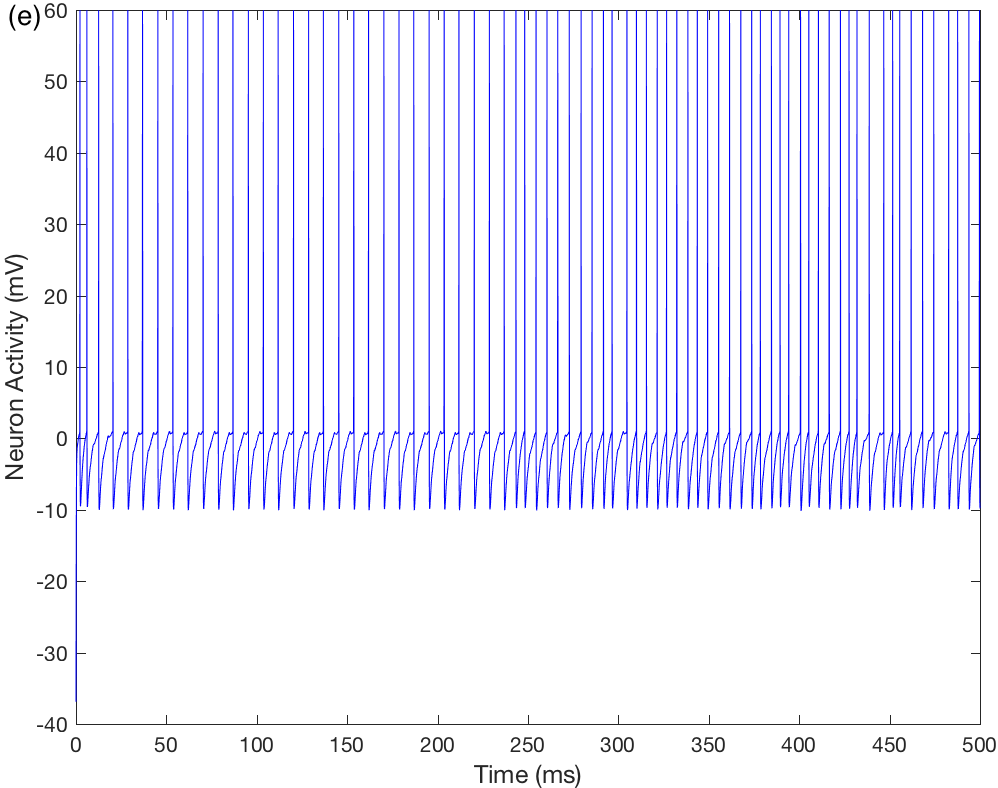}\label{fig:Tx_nic}}
{\includegraphics[width=0.495\textwidth]{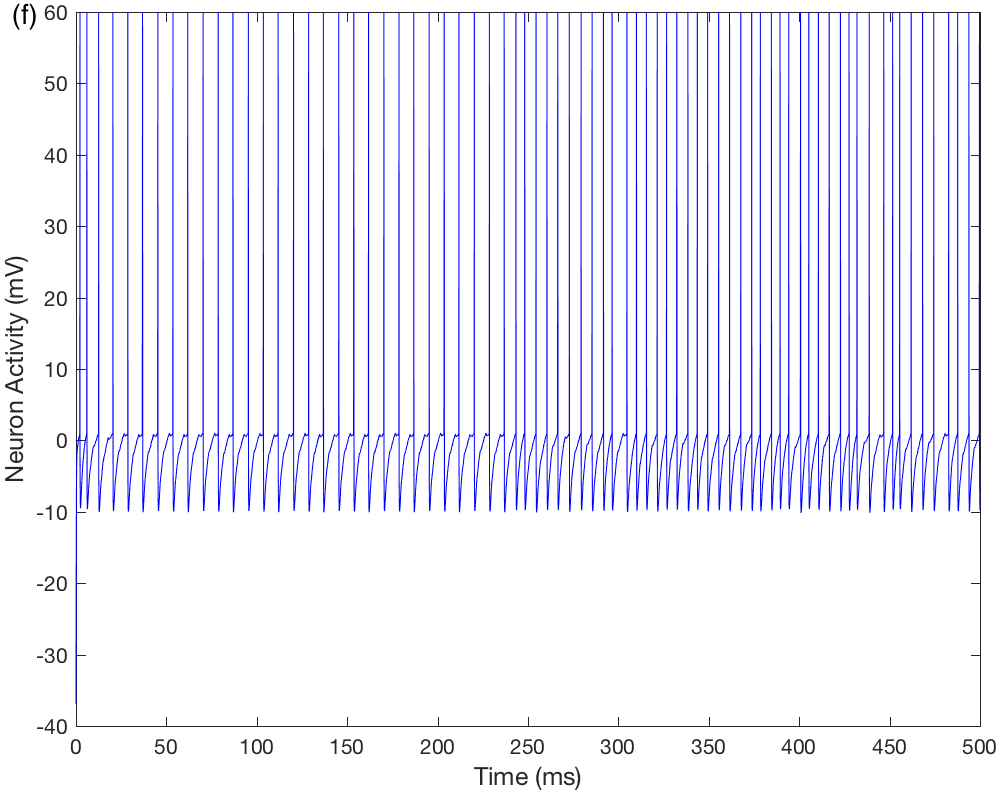}\label{fig:Ty_nic}}
\caption{\textbf{Behavior of network neurons for the network exposed to nicotine.} (a) VTA DA; (b) SNpc; (c) $TRN_x$; (d) $TRN_y$, (e) $T_x$; (f) $T_y$. }\label{fig:SleepNic}
\end{figure}

As a result of the activation of the NMDA receptors, the calcium influx, and the calcium-activated potassium current, the VTA DA neuron starts to spike in bursts. The change in the activity of the VTA DA neuron can be seen by comparing Figure~\ref{fig:SleepNormal}a to Figure~\ref{fig:SleepNic}a.

As the activity of the VTA DA neuron increases, the excitatory dopaminergic input it sends to NAcc increases as well. NAcc, in turn, strongly inhibits SNpc. Figure \ref{fig:SleepNic}b shows the decrease in the activity of SNpc. In consequence, the thalamic neurons $TRN_{x}$ and $TRN_y$ receive a low level of dopamine and their firing behavior is enhanced in comparison to the reference network (compare Figures~\ref{fig:SleepNormal}c-d to Figures~\ref{fig:SleepNic}c-d). 

Due to the mechanism of lateral inhibition between thalamic and TRN neurons in the thalamocortical circuit, when $TRN_x$ and $TRN_y$ are strongly activated they inhibit $T_y$ and $T_x$, respectively. The latter, in turn, reduce their excitatory effect on TRN neurons. As a result, the thalamic neurons $T_x$ and $T_y$ change their dynamic state from burst to tonic mode (compare Figures~\ref{fig:SleepNormal}e-f to Figures~\ref{fig:SleepNic}e-f). This corresponds to the thalamic firing behavior observed during wakefulness and suggests an increase in sleep latency.

\subsection{Influence of alcohol}

In the same way as with nicotine, starting from the reference case alcohol is added by imposing $C_2H_6O^0 = 70$ g in~\eqref{eq:alc}, which corresponds to four drinks. Note that now $[C_{10}H_{14}N_2]^0=0$. 

The alcohol stimulus activates the network by exciting the PFC and the VTA GABA neurons. In the VTA DA neuron, there is an increase in the activity of the pacemaker current due to the influence of cortical glutamatergic and VTA GABAergic stimuli.

\begin{figure}[H]
\centering
{\includegraphics[width=0.495\textwidth]{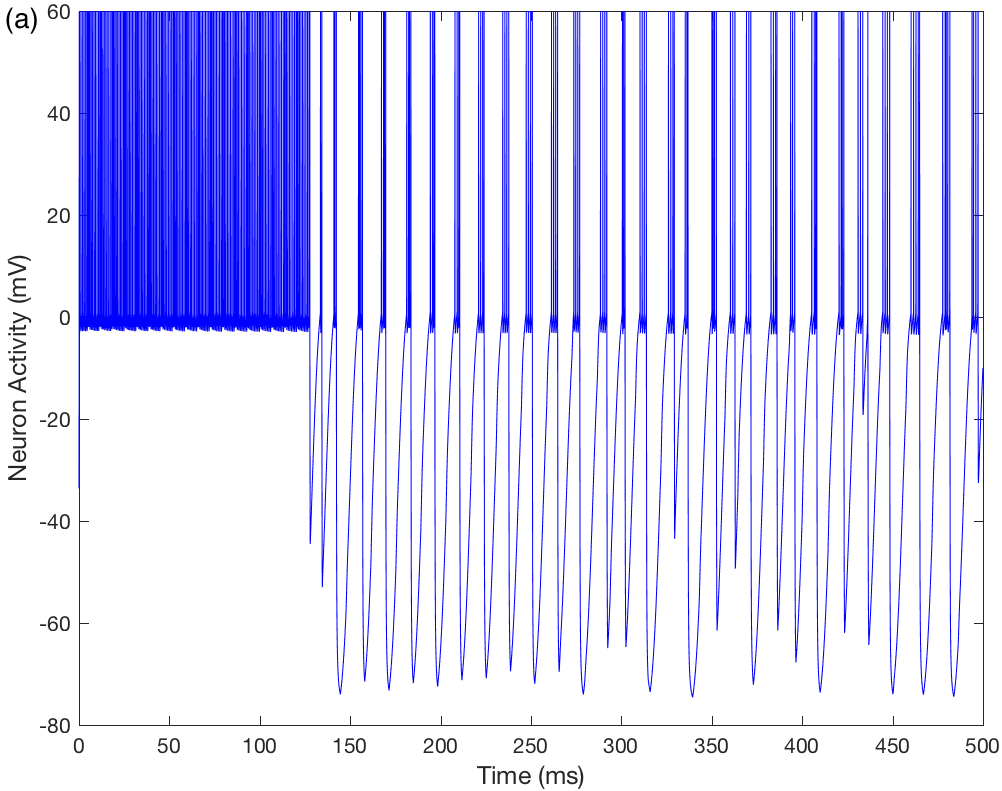}\label{fig:VTA_alcohol}}
{\includegraphics[width=0.495\textwidth]{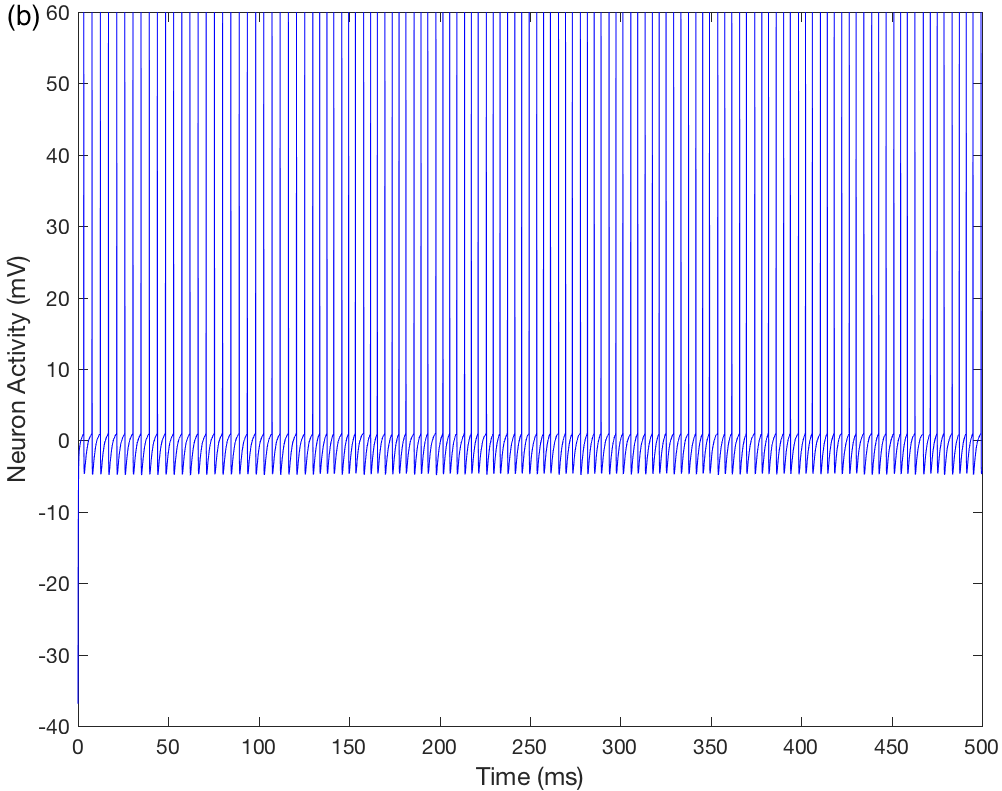}\label{fig:SN_alcohol}}
{\includegraphics[width=0.495\textwidth]{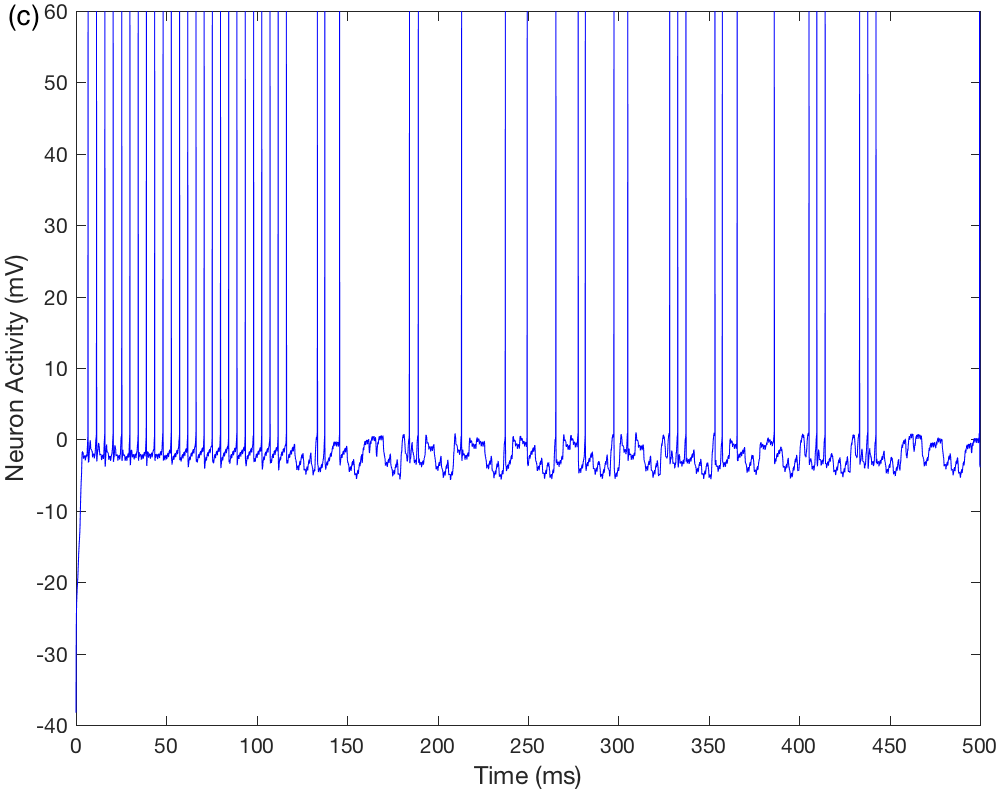}\label{fig:NRTx_alcohol}}
{\includegraphics[width=0.495\textwidth]{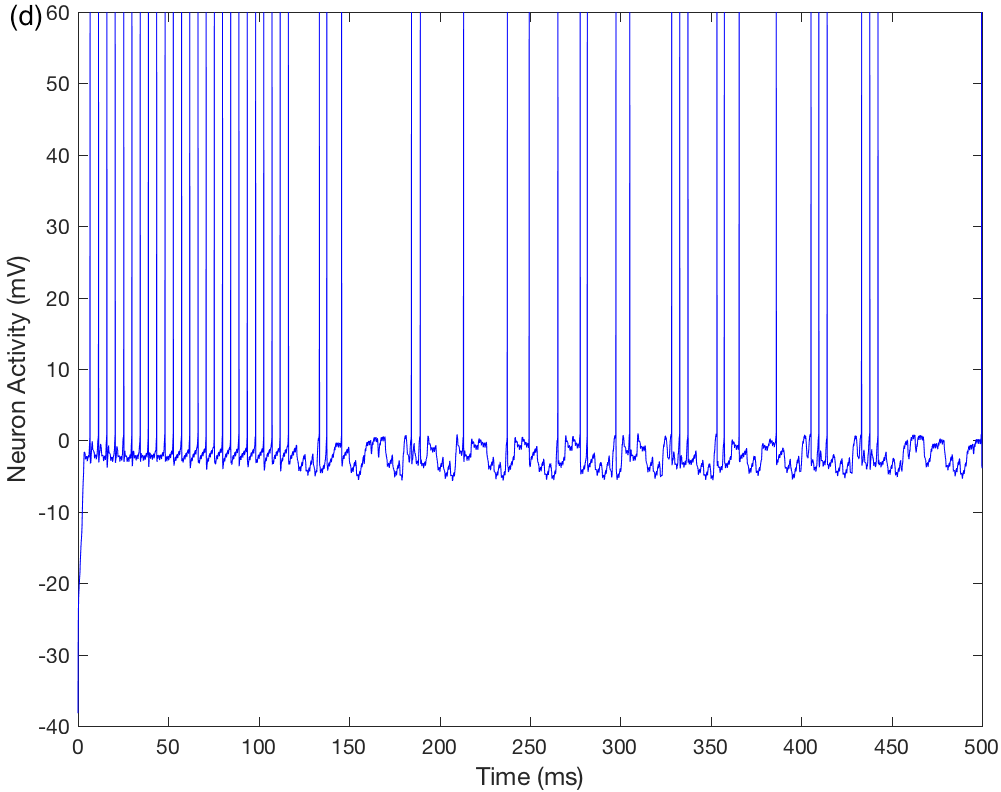}\label{fig:NRTy_alcohol}}
{\includegraphics[width=0.495\textwidth]{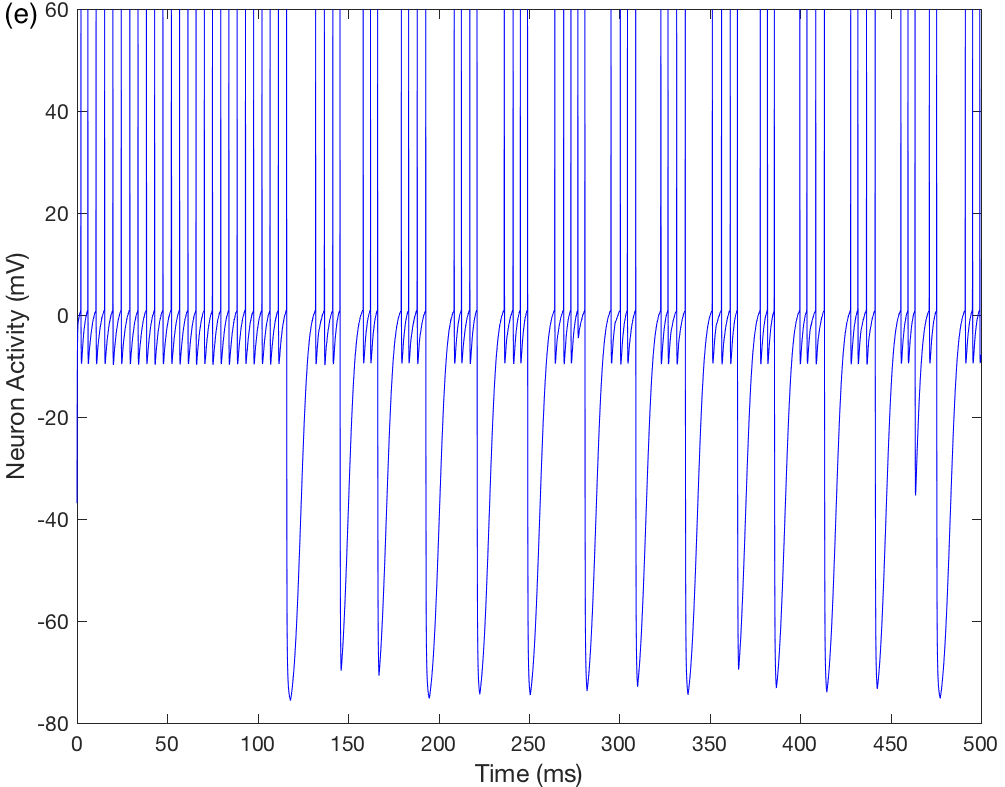}\label{fig:Tx_alcohol}}
{\includegraphics[width=0.495\textwidth]{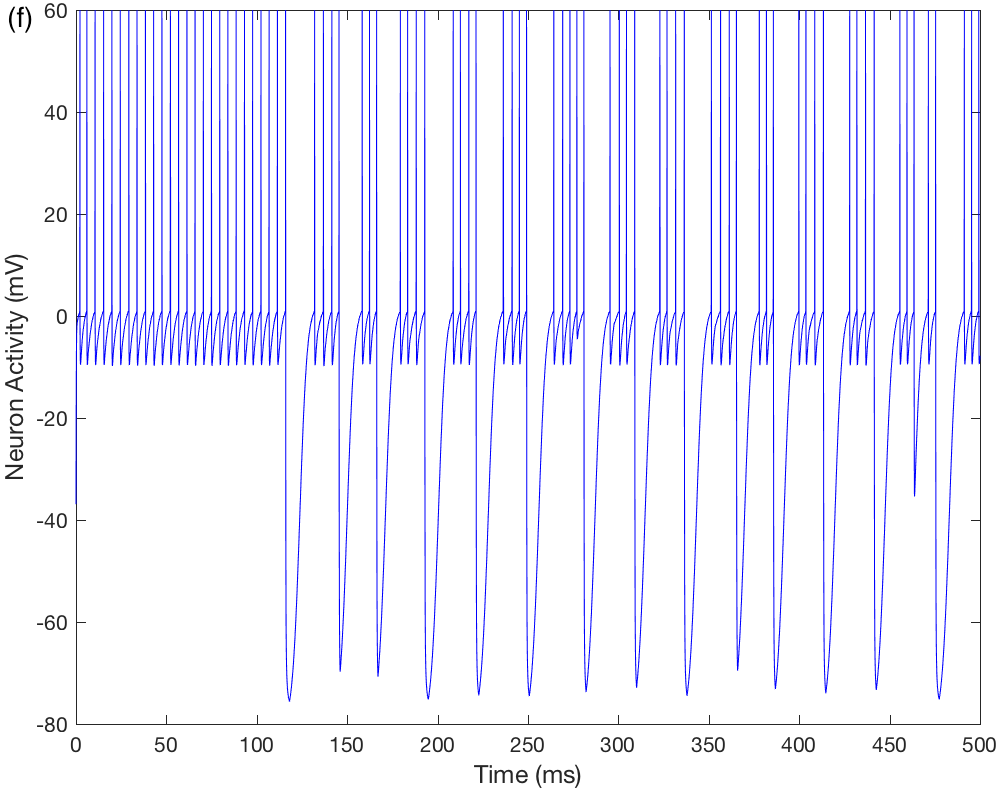}\label{fig:Ty_alcohol}}
\caption{\textbf{Behavior of network neurons for the network exposed to alcohol.} (a) VTA DA; (b) SNpc; (c) $TRN_x$; (d) $TRN_y$, (e) $T_x$; (f) $T_y$.}\label{fig:SleepAlcohol}
\end{figure}

As in the case of nicotine, the VTA DA neuron changes to burst mode because of LTP and the activation of NMDA channels (Figure \ref{fig:SleepAlcohol}a). However, alcohol inhibits the spiking of SNpr. Therefore, despite the strong inhibition of NAcc, SNpc remains active because it is no longer inhibited by the GABAergic synapse from SNpr. Figures \ref{fig:SleepAlcohol}b--d show the behavior of the SNpc, $TRN_{x}$, $TRN_{y}$, $T_x$ and $T_y$ neurons.

In this case, due to dopaminergic hyperactivity, $T_x$ and $T_y$ do not change their spiking modes. Hence, the behavior of thalamic neurons under alcohol influence does not correspond to wakefulness. Moreover, the onset of spike bursts by $T_x$ and $T_y$ is earlier under the presence of alcohol (at about 100 ms, see Figures \ref{fig:SleepAlcohol} e--f) than in comparison to the reference network, when burst firing starts at around 190 ms (Figures \ref{fig:SleepNormal} e--f). This suggests a decrease in sleep latency.

\subsection{Insomnia case}

Lesion studies have been traditionally used to assess the roles of different brain areas in sleep regulation. Insomnia and hyperactivity were observed after lesions in SNpc and SNpr~\cite{Lai1999,Gerashchenko2006}

Concerning the SNpc spiking rate, low levels of mesothalamic dopamine (dopaminergic hypoactivity) are associated with the mental rigidity observed in PD, ADHD, and ASD. On the other hand, high levels of mesothalamic dopamine (dopaminergic hyperactivity) underlie the defocusing symptoms observed in ADHD~\cite{Madureira2010,Guimaraes2017}.

\begin{figure}[H]
\centering
{\includegraphics[width=0.495\textwidth]{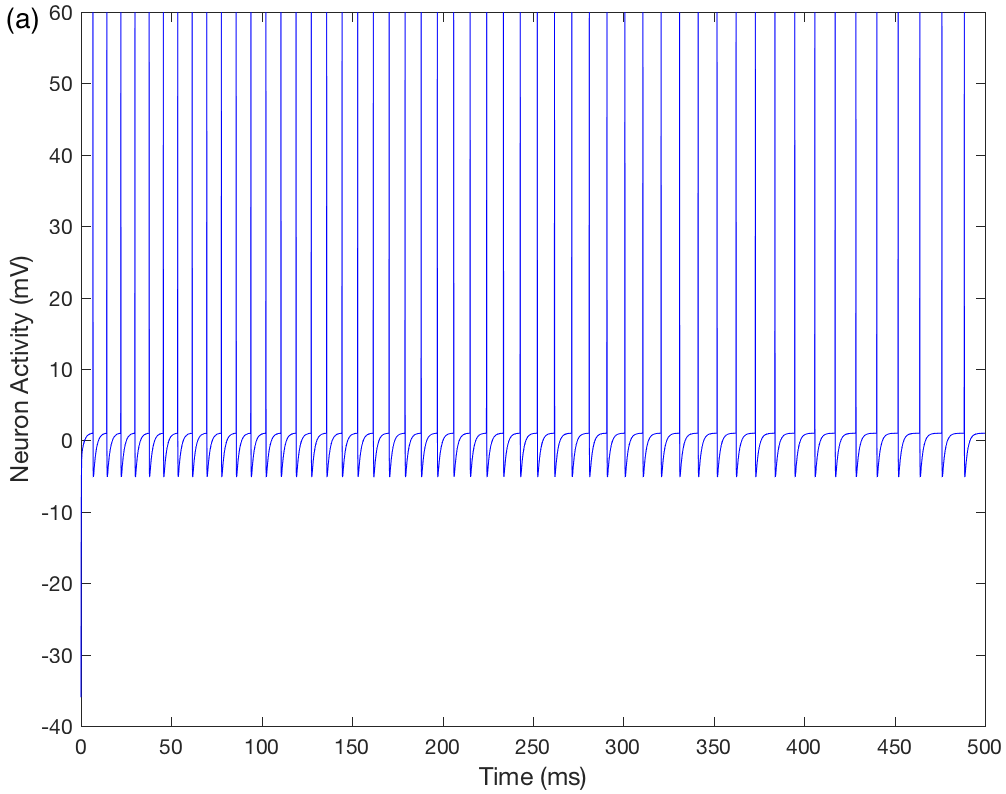}\label{fig:VTA_insomnia}}
{\includegraphics[width=0.495\textwidth]{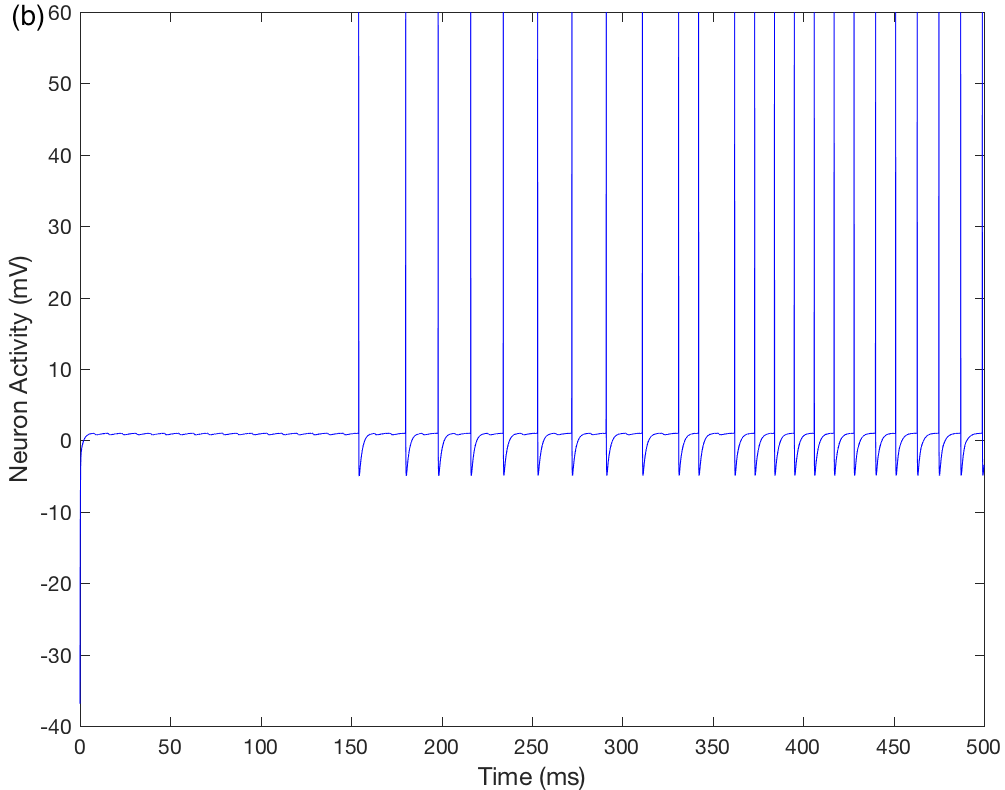}\label{fig:SN_insomnia}}
{\includegraphics[width=0.495\textwidth]{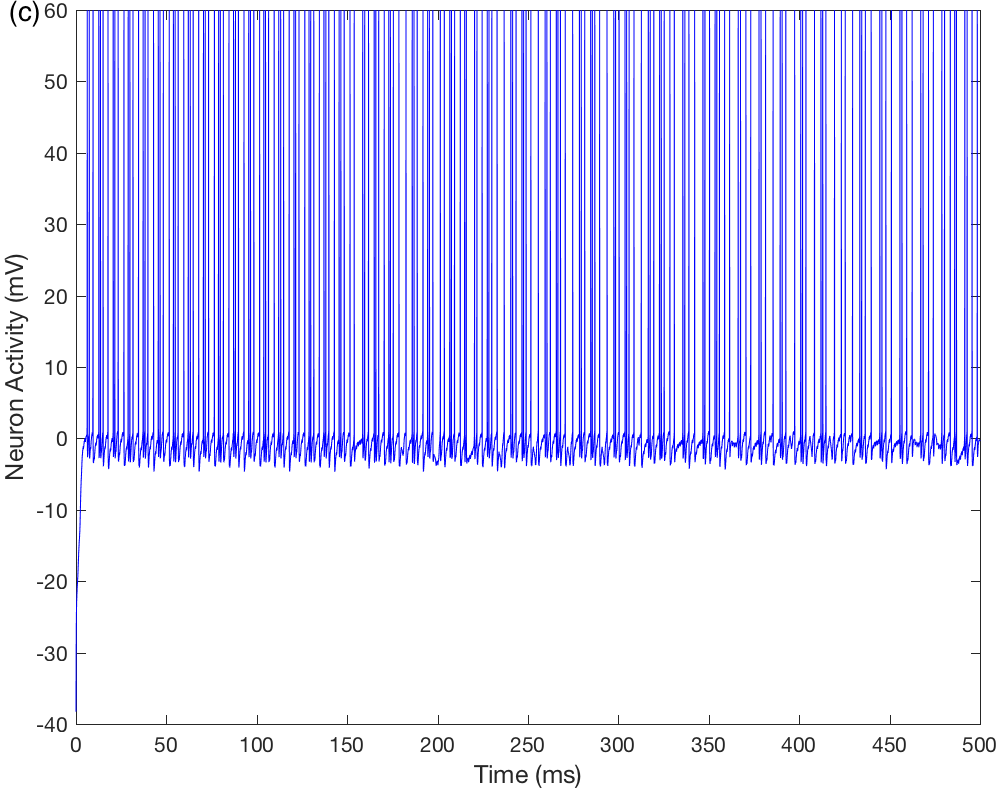}\label{fig:NRTx_insomnia}}
{\includegraphics[width=0.495\textwidth]{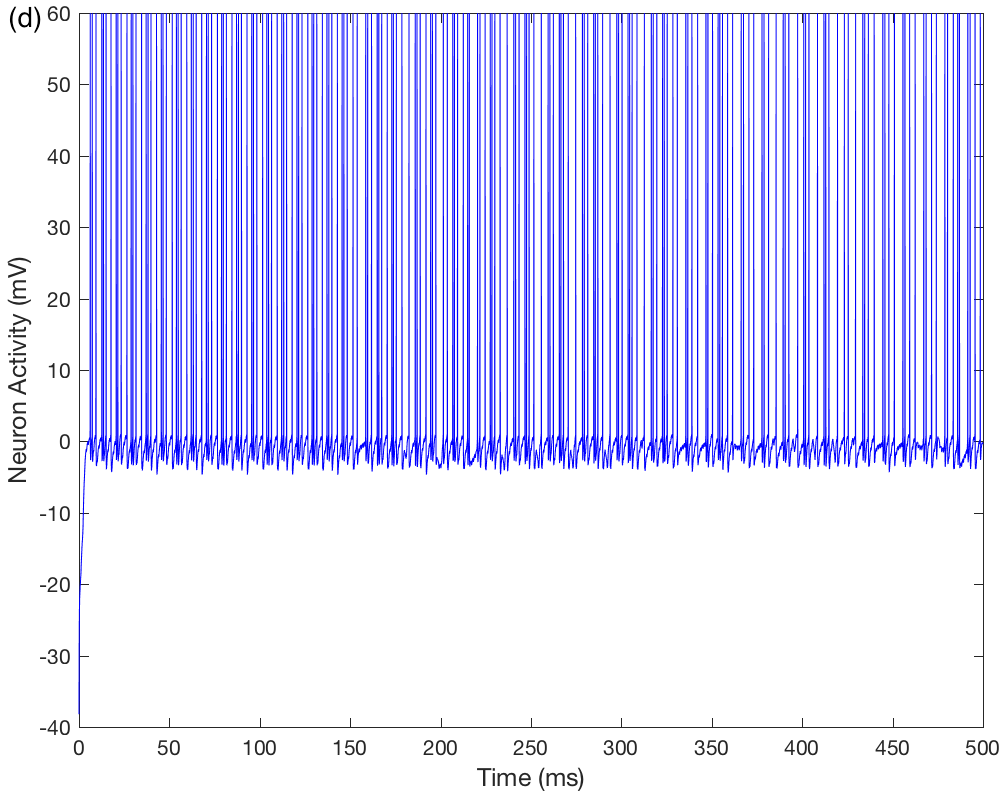}\label{fig:NRTy_insomnia}}
{\includegraphics[width=0.495\textwidth]{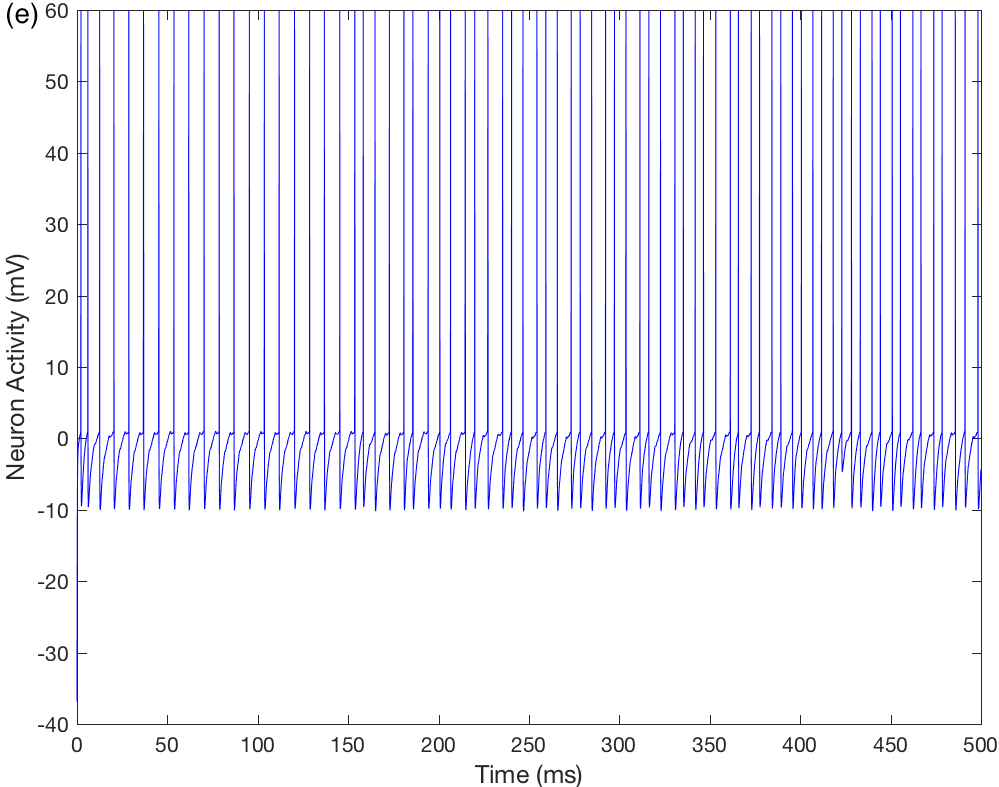}\label{fig:Tx_insomnia}}
{\includegraphics[width=0.495\textwidth]{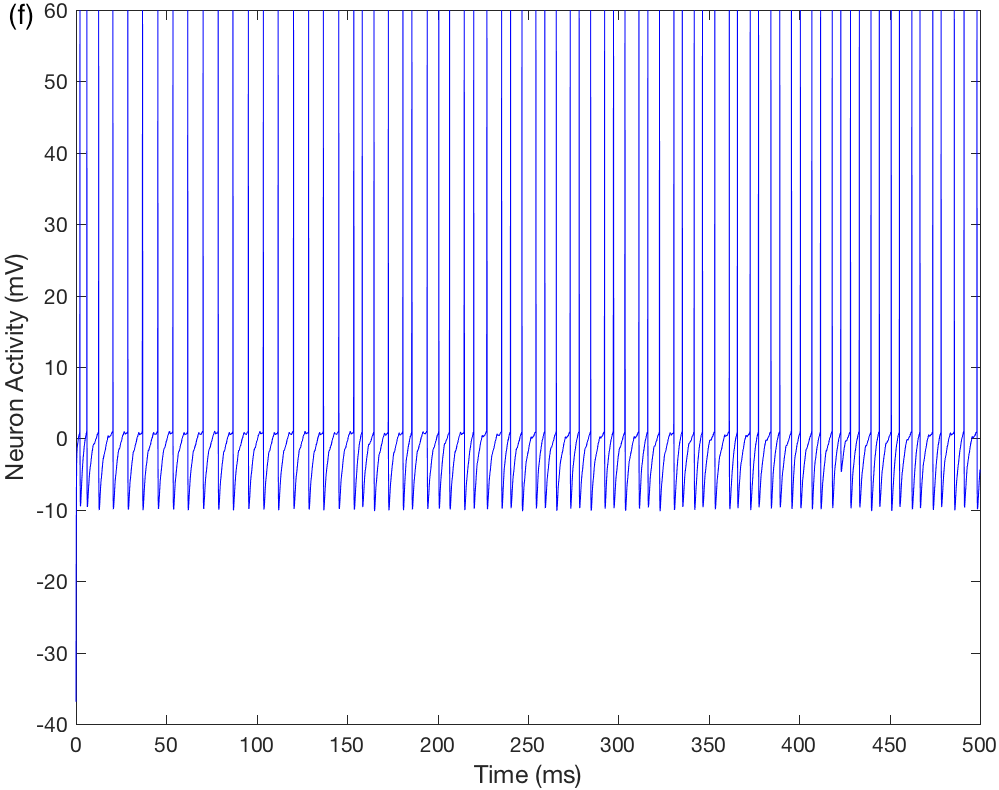}\label{fig:Ty_insomnia}}
\caption{\textbf{Behavior of network neurons under insomnia without exposure to alcohol.} (a) VTA DA; (b) SNpc; (c) $TRN_x$; (d) $TRN_y$, (e) $T_x$; (f) $T_y$.}\label{fig:SleepInsomnia}
\end{figure}

Starting from the reference network, we represented insomnia by simulating nigral dopaminergic hypoactivity (see the behavior of SNpc in Figure \ref{fig:SleepInsomnia}b). As the SNpc dopaminergic activity decreases, the $TRN_{x}$ and $TRN_{y}$ neurons become more excited (Figures \ref{fig:SleepInsomnia}c--d). Consequently, $T_x$ and $T_y$ remain firing in tonic mode as in an awake state. (Figures \ref{fig:SleepInsomnia}e--f). 

\begin{figure}[H]
\centering
{\includegraphics[width=0.495\textwidth]{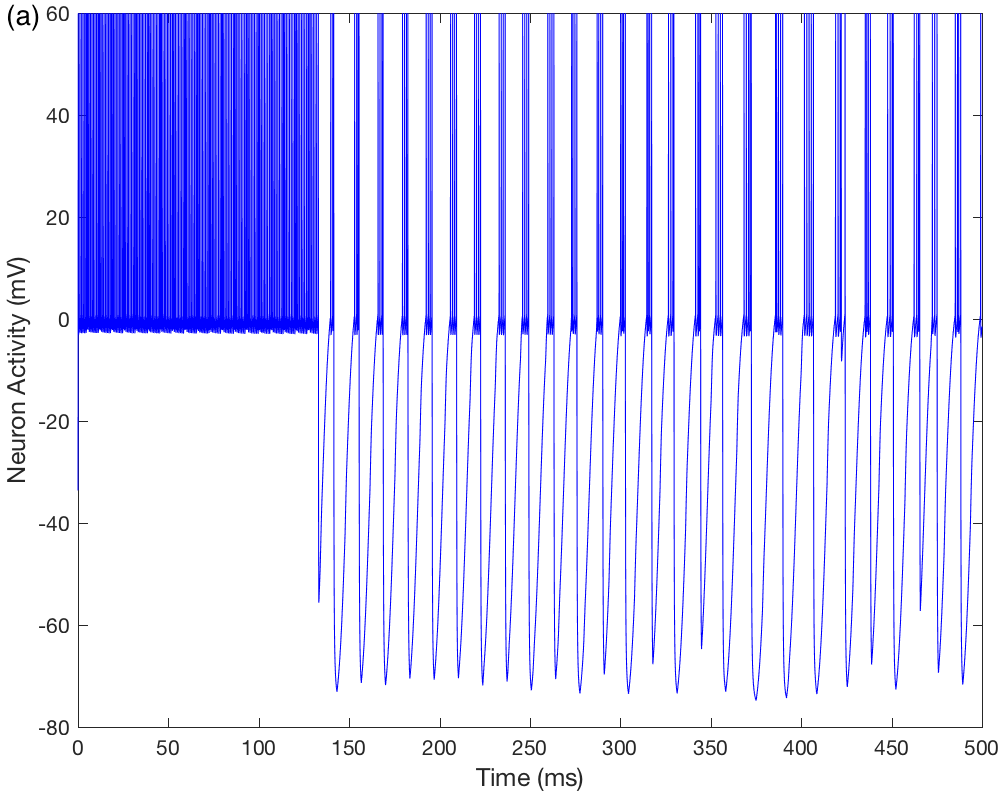}\label{fig:VTA_insomnia_alc}}
{\includegraphics[width=0.495\textwidth]{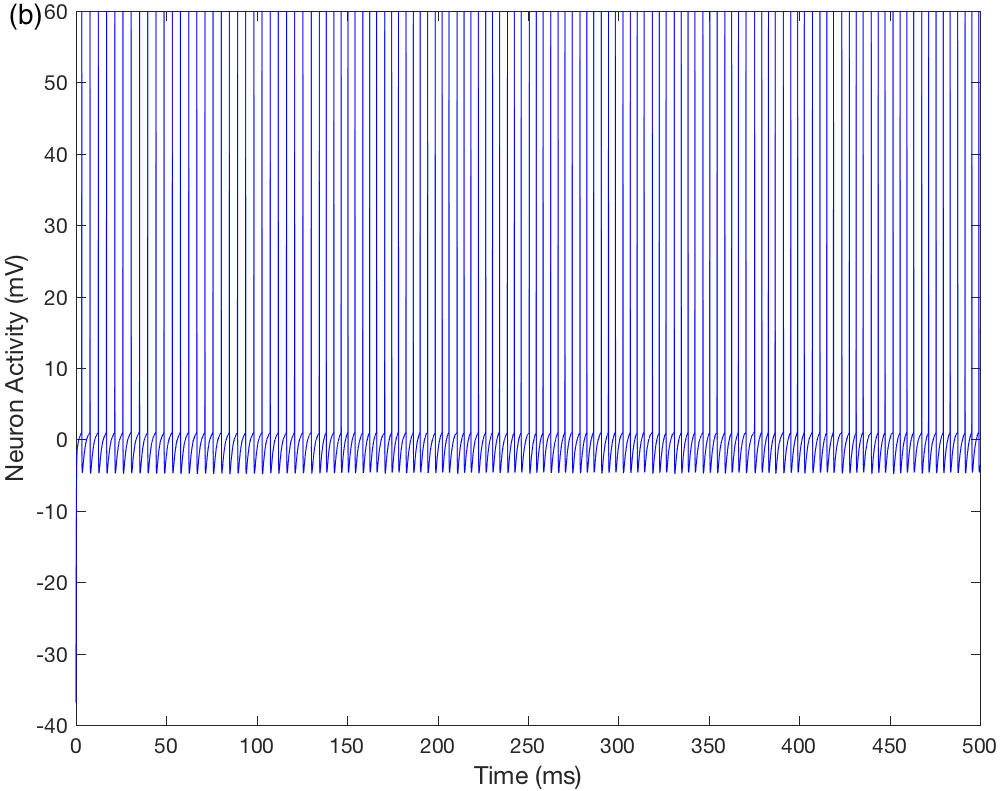}\label{fig:SN_insomnia_alc}}
{\includegraphics[width=0.495\textwidth]{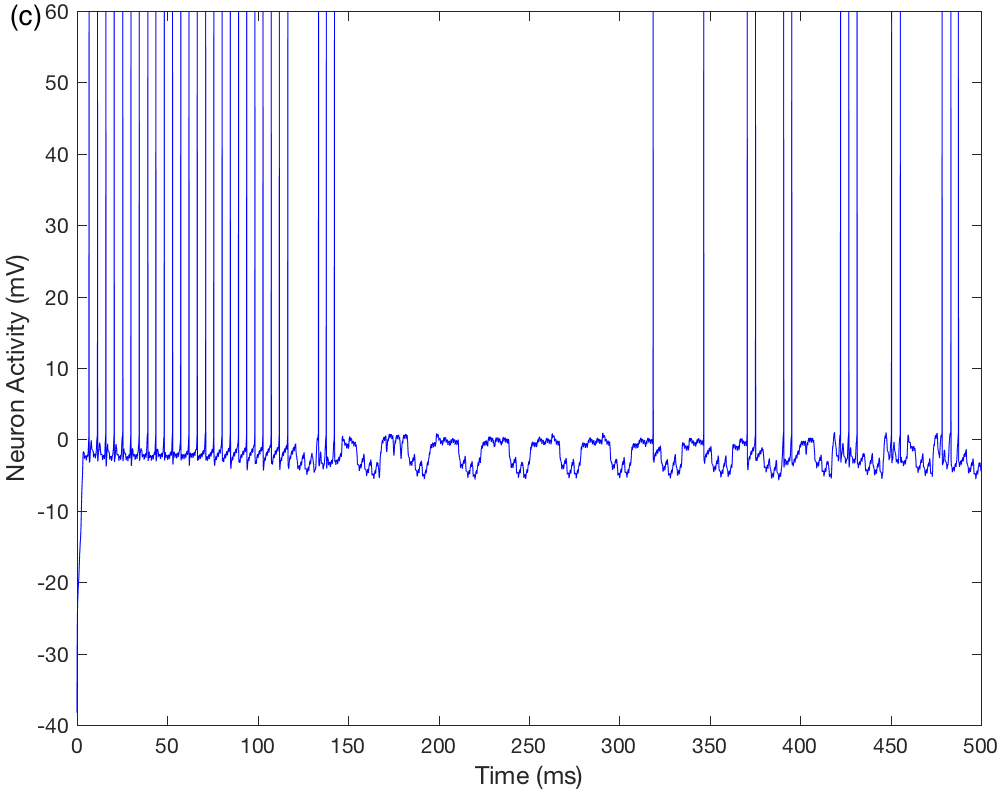}\label{fig:NRTx_insomnia_alc}}
{\includegraphics[width=0.495\textwidth]{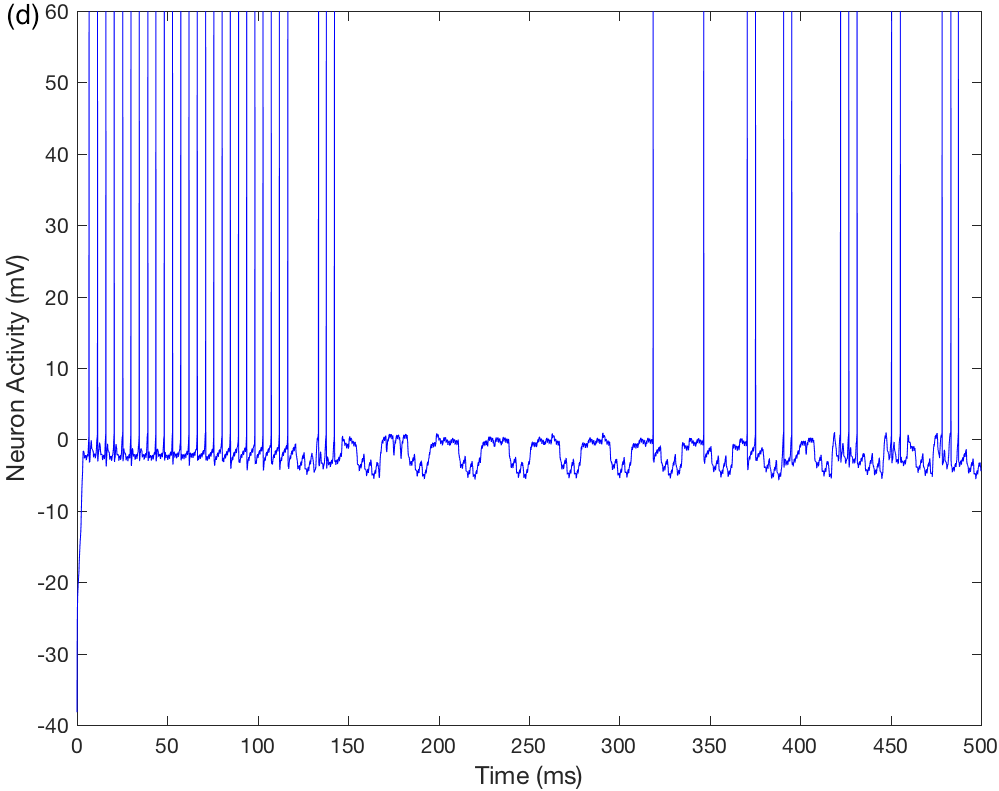}\label{fig:NRTy_insomnia_alc}}
{\includegraphics[width=0.495\textwidth]{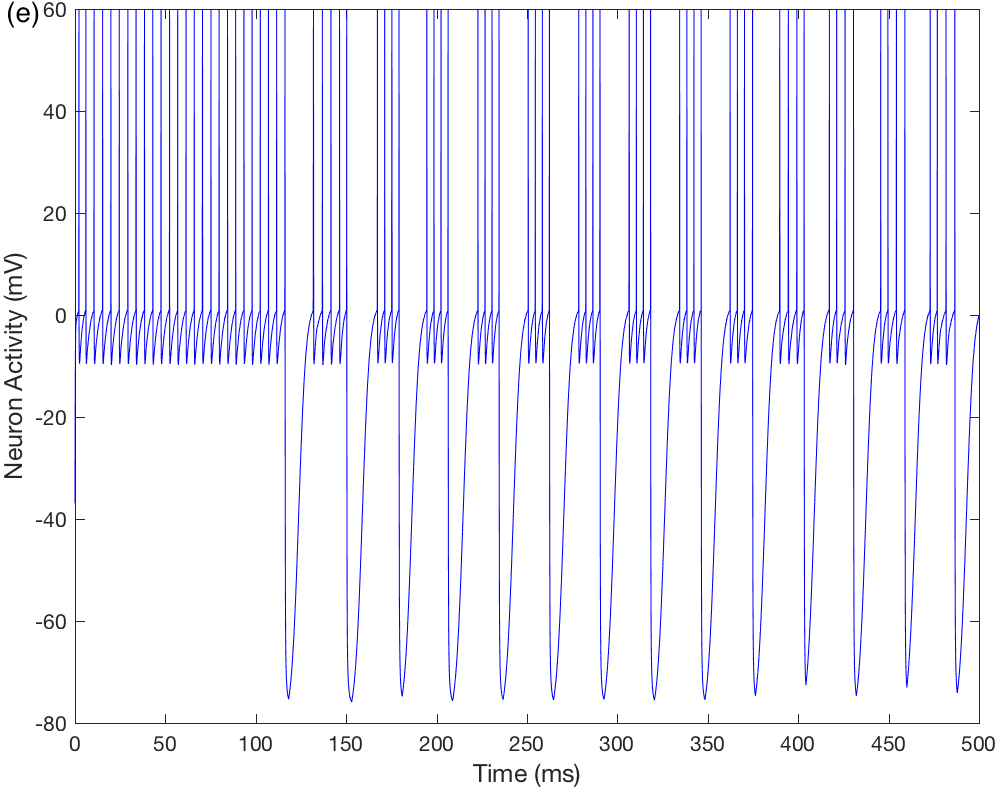}\label{fig:Tx_insomnia_alc}}
{\includegraphics[width=0.495\textwidth]{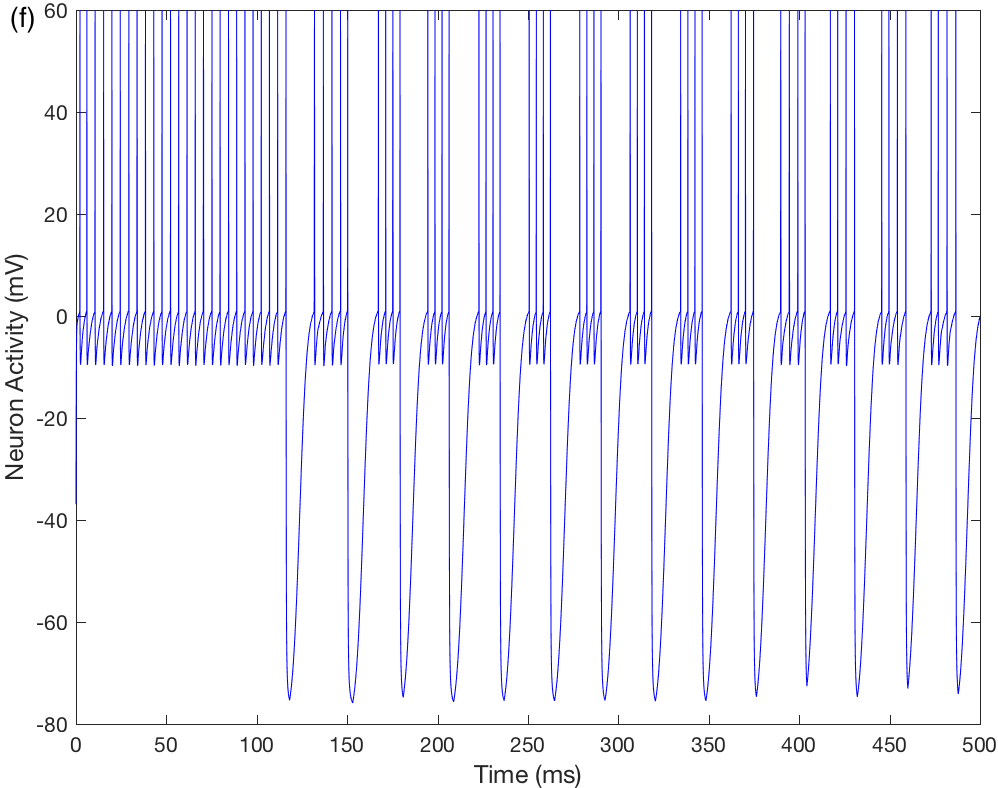}\label{fig:Ty_insomnia_alc}}
\caption{\textbf{Behavior of network neurons under insomnia with exposure to alcohol.} (a) VTA DA; (b) SNpc; (c) $TRN_x$; (d) $TRN_y$, (e) $T_x$; (f) $T_y$.}\label{fig:SleepInsomniaAlcohol}
\end{figure}

We then added alcohol to the system and the changes can be seen in Figures \ref{fig:SleepInsomniaAlcohol}a--f. Since alcohol inhibits the SNpr neuron, the activity of SNpc increases resulting in $T_x$ and $T_y$ firing as in a sleep state.

\section{Discussion}

Despite experimental evidence~\cite{Floran2004, Freeman2001}, the dopaminergic pathway from SNpc to TRN has not been extensively explored. Here, using a computational model, we explored the inhibitory modulatory role that mesothalamic dopamine has on the thalamocortical loop to study the effect of nicotine and alcohol on sleep. 

More specifically, we simulated the reward-attention circuit introduced previously by us~\cite{Guimaraes2017, Guimaraes2018} to study the effect of nicotine and alcohol on the transition from wake to sleep. Additionally, we studied how possible changes in mesothalamic dopaminergic activity may be associated with insomnia and addressed the possible effects of alcohol consumption on sleep deficit.

Many experimental studies have demonstrated that one of the determinants of sleep quality most affected by smoking is sleep latency, which increases in this case~\cite{Zhang2006, Silva2022, Jaehne2009, Manzar2017}. On the other hand, under low to moderate doses, alcohol acts as a sedative, initially promoting shortened sleep onset latency~\cite{Colrain2014, Stone1980, Stein2005}.

Starting from a circuit with dynamics that provides a predefined normal sleep state our simulations suggest that nicotine can have a stimulating effect that increases sleep latency, thus promoting wakefulness. On the other hand, the simulations indicate that alcohol can have a sedating effect that induces sleep. This result agrees with a hypothesis proposed elsewhere~\cite{Dzirasa2006}, according to which dopamine plays a role in the control of the sleep-wake cycle. Here, through a neurocomputational model, we indicate possible ways by which sleep-related states could emerge as a consequence of alterations in the mesothalamic dopamine activity. Therefore, our theoretical results give support to experimental results on the role of nicotine and alcohol on sleep~\cite{Manzar2017, Colrain2014}.

In previous work, we showed that a consequence of the decrease in the level of mesothalamic dopamine is a rigidity of attention focus causing a lack of cognitive flexibility in PD, ADHD, and ASD, suggesting a common mechanism for the lack of attention in these disorders~\cite{Madureira2010, Guimaraes2017, Guimaraes2018}. Besides, insomnia and hyperactivity were observed after lesions in SNpc and SNpr~\cite{Lai1999, Gerashchenko2006}. Since insomnia is a common symptom of these disorders, to validate the model we simulated insomnia by decreasing the level of dopamine released by the SNpc, which occurs in hyperactivity (wake) and insomnia (sleeplessness). Addition of alcohol to the system showed that alcohol has a sedative effect in situations of sleep deprivation, in line with previous findings~\cite{Colrain2014, Roehrs1999}.

Assuming the hypothesis that there is a common mechanism for the lack of attention in PD, ADHD, and ASD, we speculate that the activity alterations indicated by our simulations could also be present in the sleep deficits reported in these disorders. However, to verify the relevance of these ideas, future experiments would need to be carried out.

To summarize, this work was conducted under the hypothesis that the action of mesothalamic dopamine in the thalamocortical circuit contributes to changes in sleep latency. We studied the mediation that nicotine and alcohol have on subcortical sleep mechanisms through the action of these substances in the reward circuit and the corresponding consequences in the thalamocortical loop via \emph{substantia nigra}. Together, our results suggest insights into the distributed processing that underlies relationships between alterations in mesothalamic dopamine activity and insomnia symptoms. Our simulations suggest that both dopaminergic hypo- and hyperactivity modify thalamocortical circuit patterns and may contribute to the appearance of sleep disorders. Also, they suggest a common neuronal mechanism underlying sleep deficits in PD, ADHD, and ASD.

\section*{Appendix A}
\label{S1_Appendix}
Table~\ref{tab:parameters} presents a glossary of the parameters used in the reward-attention circuit model.

\begin{table}[ht]
{\footnotesize
\centering
\begin{tabular}{lll} 
\hline
\bf  Parameter   & \bf Description   &\bf Value/Unit  \\   
\hline
$C_i$      							  					& Membrane's capacitance                                     							&  1 $\mu$F.cm$^{-2}$  \\ 
$E_{\text{\tiny{K}}}$          						&  Reversal potential of K$^+$                   			  							&  -80 mV       \\ 
$E_{\text{\tiny{L}}}$ 	      						& Reversal potential of I$_{\text {\tiny L}}$			  								& 0 mV       \\
$\beta_{\text{\tiny{K}}}$    						& Increase rate of $g_{\text {\tiny K}}$  		 		  								& 150 \\
$\tau_{\text{\tiny{K}}}$     					    & Time constant of $g_{\text {\tiny K}}$       		  								& 1.5 ms \\
$g_{\text{\tiny{L}}}$           						& Conductance of I$_{\text {\tiny L}}$                	  								&  10 m.mhos.cm$^{-2}$    \\
$\theta$  							   					& Threshold for sodium channel's opening			  							& 1 mV \\
$M_1$          						   						& Nicotine decay rate                    						  								& 0.0001  \\ 
$M_2$          						   						& Alcohol decay rate                    						  								& 0.001  \\ 
$\mathcal{T}_{max}$	       						& Maximum concentration of neurotransmitters in the synaptic cleft & 1 mM        \\
$E_{\text{\tiny{NMDA}}}$    						& Reversal potential of $I_{\text{\tiny{NMDA}}}$    								& 0 mV \\
$E_{\text{c}}$    				   						& Reversal potential of Ca$^{++}$  						  							& 70 mV \\
$\bar{g}_{\text{c}}$ 			   					& Increase rate $g_{\text{c}}$  								  						&  1 \\
$\beta_{\text{\tiny{[Ca]}}}$ 						& Variation rate of calcium concentration 			  								& 100 \\
$\tau_{\text{\tiny{[Ca]}}}$   						& Time constant of calcium's pump 					  								& 500 ms \\
$\theta_{\text{\tiny{[Ca]}}}$						& Threshold to activate $I_{\text{\tiny{ahp}}}$ 	  								& 0.4 mV \\
$\beta_{\text{\tiny{ahp}}}$ 						& Increase rate of $g_{\text {\tiny{ahp}}}$  			  							& 100 \\
$\tau_{\text{\tiny{ahp}}}$   						& Time constant of $g_{\text{\tiny{ahp}}}$  			  							& 2 ms \\
$g_{\text{pm}}$ 				   						& Conductance of pacemaker current  					  							& 0.29 m.mhos.cm$^{-2}$ \\
$E_{\text{pm}}$ 				   						& Reversal potential of   																	& 40 mV \\
$E_{\text{sin}}^{\text{e}}$   						& Reversal potential of excitatory synapses 			   							&  40 mV  \\ 
$E_{\text{sin}}^{\text{i}}$                           & Reversal potential of inhibitory synapses  	   								    &  -40 mV       \\ 
$\hat{g}_{\text{sin}}^{\text{c-gvta}}$          & Maximal conductance of the synaptic projection cortex-gtva         &  0.18 m.mhos.cm$^{-2}$\\
$\hat{g}_{\text{sin}}^{\text{c-dvta}}$          & Maximal conductance of the synaptic projection cortex-dtva         &  1.3 m.mhos.cm$^{-2}$\\
$\hat{g}_{\text{sin}}^{\text{gvta-dvta}}$     & Maximal conductance of the synaptic projection gtva-dtva 	        & 0.3 m.mhos.cm$^{-2}$\\
$\hat{g}_{\text{sin}}^{\text{dvta-nacc}}$    & Maximal conductance of the synaptic projection dtva-nacc            &  0.5 m.mhos.cm$^-2$ \\
$\hat{g}_{\text{sin}}^{\text{nacc-snc}}$     & Maximal conductance of the synaptic projection nacc-snc              & 0.3 m.mhos.cm$^{-2}$ \\
$\hat{g}_{\text{sin}}^{\text{ppt-snc}}$       & Maximal conductance of the synaptic projection ppn-snc        		 & 0.2 m.mhos.cm$^{-2}$ \\
$\hat{g}_{\text{sin}}^{\text{snr-snc}}$       & Maximal conductance of the synaptic projection snr-snc        		 & 0.2 m.mhos.cm$^{-2}$ \\
$\hat{g}_{\text{sin}}^{\text{tc-t}}$             & Maximal conductance of the synaptic projection cortex-thalamus   & 0.1 m.mhos.cm$^{-2}$ \\
$\hat{g}_{\text{sin}}^{\text{t-nrt}}$           & Maximal conductance of the synaptic projection thalamus-trn	     & 1.3 m.mhos.cm$^{-2}$  \\
$\hat{g}_{\text{sin}}^{\text{tc-nrt}}$         & Maximal conductance of the synaptic projection cortex-trn              & 1.3 m.mhos.cm$^{-2}$ \\
$\hat{g}_{\text{sin}}^{\text{nrt-t}}$           & Maximal conductance of the synaptic projection trn-thalamus          &  0.3 m.mhos.cm$^{-2}$ \\
$\hat{g}_{\text{sin}}^{\text{ee-t}}$           & Maximal conductance of the synaptic projection stimulus-thalamus & 0.1 m.mhos.cm$^{-2}$  \\
$t_{\text{pe}}$                                        & Peak time of excitatory synaptic alpha function 								  & 1.5 ms \\
$t_{\text{pi}}$                                         & Peak time of excitatory synaptic alpha function 								  & 1.5 ms \\
$\hat{g}_{\text{c}}$                                 & Constant regulating increase $g_{\text{k-c}}$    								  &  0.4 \\
$\alpha$                                                 & Constant regulating the sigmoid inclination    								      &  1 \\ 
$\hat{g}_{\text{\tiny{d4}}}$                      & Proportion constant of dopaminergic projection 							      & 1 m.mhos.cm$^{-2}$ \\
$t_{\text{pd}}$                                        & Peak time of D$_4$ receptor's 														  & 2 ms \\
\hline
\end{tabular}
\caption{Glossary of parameters} 
\label{tab:parameters}
}
\end{table}

\section*{Acknowledgements}

\noindent KG was funded by Conselho Nacional de Desenvolvimento Cient\'{i}fico e Tecnol\'{o}gico-CNPq grant 401321/2017-7.

\section*{Additional information}

\noindent {\bf{Competing interests}} The author declare no competing interests.



\end{document}